\documentclass[a4paper,times,3p,twocolumn]{elsarticle}
\pdfoutput=1
\usepackage[utf8]{inputenc}

\usepackage[T1]{fontenc}

\usepackage{microtype}

\usepackage[x11names]{xcolor}

\usepackage{siunitx}

\usepackage{booktabs}
\usepackage{array}

\usepackage{amsmath}
\usepackage{amssymb}
\usepackage{amsthm}
\usepackage{bm}
\usepackage[hidelinks]{hyperref}

\usepackage{outlines}

\usepackage{graphicx}
\usepackage{pgf}
\usepackage{import}

\usepackage{rotating}

\usepackage{longtable}

\usepackage{algorithm}
\usepackage{algorithmic}

\usepackage{mathtools}

\RequirePackage{etoolbox}
\DeclarePairedDelimiterX\norm[1]{\lVert}{\rVert}{
\ifblank{#1}{\:\cdot\:}{#1}
}
\DeclarePairedDelimiterX\seminorm[1]{\lvert}{\rvert}{
\ifblank{#1}{\:\cdot\:}{#1}
}
\DeclarePairedDelimiterX\avg[1]{\langle}{\rangle}{
\ifblank{#1}{\:\cdot\:}{#1}
}
\DeclarePairedDelimiterX\ceil[1]{\lceil}{\rceil}{
\ifblank{#1}{\:\cdot\:}{#1}
}
\DeclarePairedDelimiterX\sprod[2]{(}{)}{
\ifblank{#1}{\cdot}{#1},\ifblank{#2}{\cdot}{#2}
}
\DeclarePairedDelimiterX\dprod[2]{\langle}{\rangle}{
\ifblank{#1}{\cdot}{#1},\ifblank{#2}{\cdot}{#2}
}

\DeclarePairedDelimiterX\set[2]{\{}{\}}
  {#1 \mathrel{}\mathclose{}\delimsize|\mathopen{}\mathrel{} #2}

\DeclareMathOperator{\optr}{tr}

\DeclareMathOperator{\opad}{ad}

\DeclareMathOperator{\arsinh}{arsinh}

\newcommand{\opre}{\operatorname{Re}}

\usepackage[font=footnotesize,labelfont=bf]{caption}
\usepackage[font=footnotesize,labelfont=bf]{subcaption}
\usepackage{multirow}

\usepackage{pxfonts}

\usepackage{import}

\usepackage{tikz}
\usepackage{tikz-3dplot}
\usetikzlibrary{3d,calc}
\usetikzlibrary{arrows.meta,backgrounds,fit,positioning,petri}
\usetikzlibrary{shapes.geometric}

\usepackage[textsize=scriptsize]{todonotes}
\usepackage{setspace}

\theoremstyle{remark}

\def\statement{\begin{minipage}[t]{.75\textwidth}
       Preprint
       \end{minipage}}

\makeatletter
\def\ps@pprintTitle{%
     \let\@oddhead\@empty
     \let\@evenhead\@empty
     \def\@oddfoot{\footnotesize\itshape
       \statement\hfill\today}%
     \let\@evenfoot\@oddfoot}
\makeatother

\journal{}

\begin{document}
\begin{frontmatter}

\title{An Eigenvalue-Free Implementation of the Log-Conformation Formulation}
\author[1]{Florian Becker\corref{cor1}}
\ead{f.becker@dlr.de}
\author[1]{Katharina Rauthmann}
\ead{katharina.rauthmann@dlr.de}
\author[2]{Lutz Pauli}
\ead{l.pauli@magmasoft.de}
\author[1]{Philipp Knechtges}
\ead{philipp.knechtges@dlr.de}
\cortext[cor1]{Corresponding author}
\address[1]{German Aerospace Center (DLR), Institute for Software Technology, High-Performance Computing, Cologne, Germany}
\address[2]{MAGMA Gießereitechnologie GmbH, Aachen, Germany}

\begin{abstract}
The log-conformation formulation, although highly successful, was from the beginning formulated as a partial differential equation that contains an, for PDEs unusual, eigenvalue decomposition of the unknown field. To this day, most numerical implementations have been based on this or a similar eigenvalue decomposition, with Knechtges et al. (2014) being the only notable exception for two-dimensional flows.

In this paper, we present an eigenvalue-free algorithm to compute the constitutive equation of the log-conformation formulation that works for two- and three-dimensional flows. Therefore, we first prove that the challenging terms in the constitutive equations are representable as a matrix function of a slightly modified matrix of the log-conformation field. We give a proof of equivalence of this term to the more common log-conformation formulations. Based on this formulation, we develop an eigenvalue-free algorithm to evaluate this matrix function. The resulting full formulation is first discretized using a finite volume method, and then tested on the confined cylinder and sedimenting sphere benchmarks.
\end{abstract}

\end{frontmatter}

\section{Introduction}\label{sec:introduction}
Since its inception~\cite{Fattal2004}, the log-conformation formulation undoubtedly has been a huge success. It had a considerable impact on attacking the High Weissenberg Number Problem (HWNP) that had riddled simulation results the decades before.

The general idea of the log-conformation formulation is simple: The conformation tensor $\mathbf{C}(x,t)\in\mathbb{R}^{d\times d}$, which, for a given instant of space $x\in\mathbb{R}^d$ and time $t$, essentially encodes a macroscopically averaged covariance of the microscopic configuration, is replaced by its matrix logarithm $\mathbf{\Psi}$ such that the conformation tensor can be recovered by the matrix exponential $\mathbf{C} = \exp \mathbf{\Psi}$. The initial motivation was to better resolve exponential stress profiles. However, another important fact is that the matrix exponential function ensures that $\mathbf{C}$ stays a symmetric positive definite matrix; a property all non-degenerate covariance matrices share. In fact, it was already known before~\cite{Hulsen1990} that a substantial class of macroscopic models respect this microscopic property also in the macroscopic equations, and the divergence of numerical simulations quite often coincided with the loss of this property.

This introduction, so far, suggests that the log-conformation formulation is a rather technical trick to enforce positivity, but in order to shed more light on the failure mechanism of numerical simulations, we want to also highlight the fact that $\mathbf{\Psi}$ naturally appears in the free energy density. E.g., in the Oldroyd-B model or Giesekus model with polymeric viscosity $\mu_P$ and relaxation time $\lambda$, it has been known for quite some time~\cite{Sarti1973,Booij1984,Grmela1987,Wapperom1998}, that the free energy density of the polymeric part $\mathcal{F}_P$ is given by $\mathcal{F}_P = \mu_P/(2\lambda)\left(\optr(\mathbf{C}) - \log\det \mathbf{C} - d\right)$. Acknowledging that $\log\det \mathbf{C} = \optr \log \mathbf{C}$ this can be rewritten in $\mathbf{\Psi}$
\begin{gather}
    \mathcal{F}_P = \frac{\mu_P}{2\lambda} \optr\left(e^\mathbf{\Psi} - \mathbf{\Psi}-\mathbf{1}\right).
\end{gather}
The implications of this statement are quite remarkable, since the second law of thermodynamics states that the free energy in total and in absence of external forces has to be non-increasing, which thus puts severe bounds on $\mathbf{\Psi}$. At best, any reasonable numerical simulation should respect this dissipative nature of the free energy, and in the light of this insight it does not seem too unexpected that it is of course easier to construct such a dissipative scheme in $\mathbf{\Psi}$ than in $\mathbf{C}$.

However, even potentially violating this physical principle does not directly explain the failure of numerical simulations. That the free energy relates to the stability of the numerical schemes is mostly an indication from the known mathematical existence results in the discretized setting~\cite{Jourdain2006,Hu2007,Boyaval2009,Knechtges2018}. They all use the free energy to prove existence, and it is thus not unreasonable to conclude that the free-energy-dissipative nature of a numerical scheme and the existence of a numerical solution essentially appear as two sides of the same medal. It is this insight that brings us to the conclusion that the log-conformation formulation has, as far as the fully nonlinear numerical schemes are concerned, solved the HWNP.

Nonetheless, all these advantages have a drawback: the resulting constitutive equation as formulated in $\mathbf{\Psi}$ becomes much more complex. Beginning from the first log-conformation formulation, almost all new constitutive equations in $\mathbf{\Psi}$ made use of an eigenvalue decomposition of $\mathbf{\Psi}$. The latter is highly unusual for a partial differential equation in the sense that the new equation contains an eigenvalue decomposition of the unknown degrees of freedom. Two notable exceptions to this were~\cite{Knechtges2014}, which introduced an eigenvalue-free formulation in two-dimensions, and~\cite{Knechtges2015}, which substituted the eigenvalue-based terms by a Cauchy-type integral in the three-dimensional setting. However, \cite{Knechtges2015} still relied on eigenvalues for the actual numerical computation, since Cauchy integrals are known to be prone to numerical cancellation issues. With this paper we will bridge the gap, and provide an eigenvalue-free implementation also for the three-dimensional setting.

In order to derive this new algorithm, we will use a formulation of the constitutive equation that was introduced in~\cite{Knechtges2018}. Since we do not want to derive a constitutive equation from first principles, as it was done in~\cite{Knechtges2018}, and for the sake of brevity, we rather make the connection to the more popular log-conformation formulations in Section~\ref{sec:theory}. There it will be shown that all these log-conformation formulations are equal in perfect arithmetic.

Given this new formulation, we will, in Section~\ref{sec:eigVal-Free_Algo}, derive an algorithm that allows for the eigenvalue-free numerical evaluation of this term. This algorithm is in principle not bound to a particular discretization scheme, and thus suitable for either finite element or finite volume discretizations.

In Section~\ref{sec:numimpl}, we then subsequently introduce shortly the finite volume aspects of the numerical scheme we chose to conduct our experiments in. Our implementation is based on the RheoTool software~\cite{rheoTool}, and since our reformulation is independent of the actual discretization of differential operators, we keep the changes minimal. Therefore, we will also not discuss matters of stable discretization of the incompressible Navier--Stokes equations using the finite volume method, and rather refer to~\cite{pimenta2017stabilization}. Furthermore, we also want to point the interested reader to the review paper~\cite{alves2021numerical} and the references therein for a broader picture on the simulation of viscoelastic fluid flows.

In Section~\ref{sec:benchmarks}, we present two benchmarks: the confined cylinder and the sedimenting sphere. Both benchmarks consider fluid flow around an obstacle, a cylinder and a sphere, respectively. Furthermore, drag coefficient values are computed and compared to results from selected publications.

At last, we also want, for the sake of completeness, mention that other schemes than the log-conformation formulation have been proposed and successfully employed to enforce the positive-definiteness of $\mathbf{C}$. Most notably are the square-root-based approach in~\cite{Balci2011} or the Cholesky-type decomposition in~\cite{Vaithianathan2003}, as well as the more recently introduced contravariant deformation tensor approach~\cite{carrozza2019viscoelastic}.

\section{Theory of Log-Conformation Formulations}\label{sec:theory}
Over the course of the years there have been many different log-conformation formulations, which in perfect arithmetic
all yield the same result. Starting point is a constitutive equation of the symmetric conformation tensor $\mathbf{C}$
\begin{gather}\label{eqn:conf}
\begin{split}
	\partial_t \mathbf{C} + (\mathbf{u} \cdot \nabla) \mathbf{C}
		- \nabla\mathbf{u}\, \mathbf{C} - \mathbf{C}\, \nabla\mathbf{u}^T
		= - P(\mathbf{C})\, .
\end{split}
\end{gather}
Here we have chosen the convention that $[\nabla\mathbf{u}]_{ij} = \partial_i u_j$ is the Jacobian of the velocity field $\mathbf{u}$, such that the left-hand side of the equation corresponds to the upper-convected derivative of the conformation tensor.
$P$ is in full generality a function of $\mathbf{C}$ that maps $\mathbf{C}$ to another symmetric matrix that commutes with $\mathbf{C}$, i.e., $P(\mathbf{C})\, \mathbf{C} = \mathbf{C}\, P(\mathbf{C})$. Common choices, that are relevant for later sections of this paper, are the Oldroyd-B model $P(\mathbf{C}) = \frac{1}{\lambda} \left(\mathbf{C} - \mathbf{1}\right)$ with a relaxation time $\lambda$, as well as the Giesekus model $P(\mathbf{C}) = \frac{1}{\lambda} \left(\mathbf{1} + \alpha \left(\mathbf{C} - \mathbf{1}\right)\right) \left(\mathbf{C} - \mathbf{1}\right)$ with an additional mobility parameter $\alpha$.

The log-conformation formulation now replaces $\mathbf{C}$ by an auxiliary symmetric tensor $\mathbf{\Psi}$ such that the two relate via the matrix exponential function $\mathbf{C} = \exp(\mathbf{\Psi})$. As stated in the introduction, the replacement has the advantage that $\mathbf{C}$ stays positive definite. However, this necessitates a new constitutive equation for $\mathbf{\Psi}$ that replaces Eq.~\eqref{eqn:conf}. In the formulation that will be used throughout this paper, this equation is stated as
\begin{gather}\label{eqn:logconf}
\begin{split}
	0 = \partial_t \mathbf{\Psi} + (\mathbf{u} \cdot \nabla) \mathbf{\Psi}
		+ \mathbf{\Psi} \omega(\mathbf{u}) - \omega(\mathbf{u}) \mathbf{\Psi}\\\qquad
		- 2\, f(\opad \mathbf{\Psi})\, \epsilon(\mathbf{u})
		+ P(e^{\mathbf{\Psi}}) e^{-\mathbf{\Psi}}\, ,
\end{split}
\end{gather}
where $\omega(\mathbf{u}) \coloneqq (\nabla \mathbf{u} - \nabla \mathbf{u}^T)/2$ is the vorticity tensor and $\epsilon(\mathbf{u}) \coloneqq (\nabla \mathbf{u} + \nabla \mathbf{u}^T)/2$ is the strain tensor.
The most important part, however, is $f(\opad \mathbf{\Psi})\, \epsilon(\mathbf{u})$, for which different formulations and numerical algorithms exist. Note that this term distinguishes the different log-conformation formulations, which in perfect arithmetic all yield the same numerical results.

The formulation chosen here is in full generality proven to be equal to the original conformation equation~\eqref{eqn:conf} in~\cite[Theorem A.42]{Knechtges2018}. We will refrain here from an exposition that shows this equivalence from first principles and in full generality. Instead, we explain our formulation first by defining $f(\opad \mathbf{\Psi})\, \epsilon(\mathbf{u})$ properly, and then show the equivalence of the different log-conformation formulation to this formulation in a second step. Those already familiar with one of the other log-conformation formulations should thus more easily grasp the formulation in Eq.~\eqref{eqn:logconf}.

For the definition, we first introduce some terminology:
Given two square matrices $\mathbf{A},\mathbf{B} \in \mathbb{R}^{d\times d}$, we define the commutator $[\mathbf{A}, \mathbf{B}] = \mathbf{A} \mathbf{B} - \mathbf{B} \mathbf{A}$. Then, the adjoint operator $\opad \mathbf{A}: \mathbb{R}^{d\times d} \to \mathbb{R}^{d\times d}$ is defined as the linear operator that maps any matrix $\mathbf{B}$ to $[\mathbf{A}, \mathbf{B}]$, i.e.,
\begin{gather*}
    \opad \mathbf{A}\, (\mathbf{B}) \coloneqq [\mathbf{A}, \mathbf{B}]\, .
\end{gather*}
The important point to note here, which will become crucial for our algorithm, is that $\opad \mathbf{A}$ is a linear operator, i.e., a homomorphism from a vector space $\mathbb{R}^{d \times d}$ to the same vector space $\mathbb{R}^{d\times d}$. As such, it is in linear algebra terms representable as a matrix: There exists a matrix $\mathbf{M}$ in $\mathbb{R}^{d^2 \times d^2}$ such that
\begin{gather*}
    \opad \mathbf{A}\, (\mathbf{B}) = \mathbf{M}\, \mathbf{\tilde{b}}\, ,
\end{gather*}
where $\mathbf{\tilde{b}}$ is just a reshaping of the matrix $\mathbf{B}$ to a vector in $\mathbb{R}^{d^2}$, and the product between $\mathbf{M}$ and $\mathbf{\tilde{b}}$ is the usual matrix vector product. To make this more explicit and less abstract, e.g., in the $d=2$ case we could write $\mathbf{D} = \opad \mathbf{A}\, (\mathbf{B}) = \mathbf{A} \mathbf{B} - \mathbf{B}\mathbf{A}$ as
\begin{gather*}
    \begin{pmatrix} D_{11}\\D_{12}\\D_{21}\\D_{22} \end{pmatrix}
    = \begin{pmatrix} 0 & -A_{21} & A_{12} & 0\\
            -A_{12} & A_{11}-A_{22} & 0 & A_{12}\\
            A_{21} & 0 & A_{22}-A_{11} & -A_{21}\\
            0 & A_{21} & -A_{12} & 0\end{pmatrix}
    \begin{pmatrix} B_{11}\\B_{12}\\B_{21}\\B_{22} \end{pmatrix}\, .
\end{gather*}
For the sake of brevity, and since it will not be used for the actual algorithm, we skip the related formula for $d=3$. For the rest of the paper $d$ will be fixed to $d=3$.

Hence, for a given instant of space $x$ and time $t$, the operation $\opad \mathbf{\Psi}\, (\epsilon(\mathbf{u}))$ can be thought of as a matrix-vector multiplication of a matrix $\mathbb{R}^{9\times 9}$ and a vector $\mathbb{R}^9$ for the three-dimensional case. In the following, as is customary for linear operators and especially matrix-vector multiplications, we will omit the parentheses around the argument and just write $\opad \mathbf{\Psi}\, \epsilon(\mathbf{u})$.

Lastly, we define $f(\opad \mathbf{\Psi})$ as the application of the function
\begin{gather}
    f(x) = \frac{x/2}{\tanh(x/2)}
\end{gather}
to the $9\times 9$-dimensional matrix that represents $\opad \mathbf{\Psi}$.

To summarize: For each instant of space and time, we think of $f(\opad \mathbf{\Psi})\, \epsilon(\mathbf{u})$ as the function $f$ applied to a $9\times 9$-matrix representation of $\opad \mathbf{\Psi}$, and the result being multiplied with a $9$-vector representation of $\epsilon(\mathbf{u})$.

This is already, modulo several optimizations for symmetric matrices, the gist of the Algorithm~\ref{alg:fopad} in the following section: We will evaluate this function $f$ of a matrix that represents $\opad \mathbf{\Psi}$ without the need to do an eigenvalue decomposition of $\mathbf{\Psi}$.

This brings us to the second part of this section: The question how previous log-conformation formulations have evaluated this term.

A straightforward way is using the Taylor expansion of $f$, which is given by
\begin{gather}\label{eqn:taylorf}
    f(x) = \sum_{n=0}^\infty \frac{B_{2n}}{(2n)!} x^{2n}\, ,
\end{gather}
where $B_{2n}$ are the even Bernoulli numbers. Substituting $x$ by $\opad\mathbf{\Psi}$ yields
\begin{align}\label{eqn:taylorfad}
\begin{split}
    f(\opad \mathbf{\Psi})\, \epsilon(\mathbf{u})
        =& \sum_{n=0}^\infty \frac{B_{2n}}{(2n)!} \opad^{2n} \mathbf{\Psi}\, \epsilon(\mathbf{u})\\
        =& \sum_{n=0}^\infty \frac{B_{2n}}{(2n)!} \underbrace{[\mathbf{\Psi},[\mathbf{\Psi}, [\ldots ,[\mathbf{\Psi}}_{\mbox{$2n$ commutators}},\epsilon(\mathbf{u})]\ldots]\, .
\end{split}
\end{align}
This formulation was first proven in~\cite[Theorem 1]{Knechtges2014}. However, as it was noted in~\cite{Knechtges2014}, this formulation alone is for practical numerical simulations not directly usable, since $f(x)$ has singularities at $\pm 2\pi i$, which limits the convergence radius of the Taylor expansion.

To make a connection with the eigenvalue-based formulations, we introduce the eigenvalue decomposition of $\mathbf{\Psi}$
\begin{gather}
    \mathbf{\Psi} = \mathbf{O} \begin{pmatrix} \lambda_1 & &\\ & \lambda_2 & \\ &&\lambda_3 \end{pmatrix}
        \mathbf{O}^T\, ,
\end{gather}
with $\mathbf{O} = \left(\mathbf{e}_1, \mathbf{e}_2, \mathbf{e}_3\right)$ being an orthogonal matrix. $\lambda_i$ are the eigenvalues and $\mathbf{e}_i$ the corresponding eigenvectors. For the following, it is also customary to introduce the projection operators $\mathbf{P}_i = \mathbf{e}_i \mathbf{e}_i^T$, which allows us to state the decomposition in the form $\mathbf{\Psi} = \sum_i \lambda_i \mathbf{P}_i$. Furthermore, the fact $\mathbf{O}\mathbf{O}^T = \mathbf{1}$ yields $\mathbf{1} = \sum_i \mathbf{P}_i$.

In combination, we can thus state
\begin{align*}
    \opad \mathbf{\Psi}\, \epsilon(\mathbf{u}) &= \mathbf{\Psi} \epsilon(\mathbf{u})
            - \epsilon(\mathbf{u}) \mathbf{\Psi} \\
        &=\sum_{i,j} (\lambda_i - \lambda_j) \mathbf{P}_i \epsilon(\mathbf{u}) \mathbf{P}_j\, .
\end{align*}
Furthermore, it is not difficult to see by algebraic manipulations that this can be generalized to any polynomial $p$
\begin{align*}
    p(\opad \mathbf{\Psi})\, \epsilon(\mathbf{u})
        &= \sum_{i,j} p(\lambda_i - \lambda_j) \mathbf{P}_i \epsilon(\mathbf{u}) \mathbf{P}_j\, .
\end{align*}
It is now mostly an application of the Stone--Weierstrass theorem that this not only holds for polynomials, but also for the continuous function~$f$
\begin{align}\label{eqn:firsteigenvallogconf}
    f(\opad \mathbf{\Psi})\, \epsilon(\mathbf{u})
        &= \sum_{i,j} f(\lambda_i - \lambda_j) \mathbf{P}_i \epsilon(\mathbf{u}) \mathbf{P}_j\, .
\end{align}
Eq.~\eqref{eqn:firsteigenvallogconf} is the formulation as it was used for numerical evaluation in~\cite{Knechtges2015,Knechtges2018,fernandes2022fully}, and is in some sense closest to what was used in~\cite{Saramito2014}.

To see that the more popular eigenvalue-based formulations are just variations of this formulation, we also need to incorporate the rotational term
\begin{align*}
    &\mathbf{\Psi} \omega(\mathbf{u}) - \omega(\mathbf{u}) \mathbf{\Psi}\\
        &\qquad= \sum_{i,j} (\lambda_i - \lambda_j) \frac{e^{\lambda_i}-e^{\lambda_j}}{e^{\lambda_i}-e^{\lambda_j}}
            \mathbf{P}_i \frac{\nabla\mathbf{u} - \nabla \mathbf{u}^T}{2} \mathbf{P}_j \, .
\end{align*}
Using $\tanh((\lambda_i-\lambda_j)/2) = (e^{\lambda_i} - e^{\lambda_j})/(e^{\lambda_i} + e^{\lambda_j})$ we can combine this with Eq.~\eqref{eqn:firsteigenvallogconf} to get
\begin{align}
\begin{split}
    &\mathbf{\Psi} \omega(\mathbf{u}) - \omega(\mathbf{u}) \mathbf{\Psi}
            - 2\, f(\opad \mathbf{\Psi})\, \epsilon(\mathbf{u})\\
        &\qquad= -\sum_{i,j} \frac{\lambda_i - \lambda_j}{e^{\lambda_i}-e^{\lambda_j}}
            \mathbf{P}_i \left(e^{\lambda_j} \nabla\mathbf{u}
                + e^{\lambda_i} \nabla \mathbf{u}^T \right)\mathbf{P}_j \, .
\end{split}
\end{align}
Furthermore, note that $\lim_{\lambda_j \to \lambda_i} \frac{\lambda_i - \lambda_j}{e^{\lambda_i}-e^{\lambda_j}} = e^{-\lambda_i}$, which allows us to split off the $i=j$ part
\begin{align}\label{eqn:sameashulsen}
\begin{split}
    &\mathbf{\Psi} \omega(\mathbf{u}) - \omega(\mathbf{u}) \mathbf{\Psi}
            - 2\, f(\opad \mathbf{\Psi})\, \epsilon(\mathbf{u})\\
        &\qquad= -2 \mathbf{B} -\sum_{i\neq j} \frac{\lambda_i - \lambda_j}{e^{\lambda_i}-e^{\lambda_j}}
            \mathbf{P}_i \left(e^{\lambda_j} \nabla\mathbf{u}
                + e^{\lambda_i} \nabla \mathbf{u}^T \right)\mathbf{P}_j \, ,
\end{split}
\end{align}
with
\begin{align}
    \mathbf{B} =& \sum_i \mathbf{P}_i\, \epsilon(\mathbf{u})\, \mathbf{P}_i
        = \sum_i \mathbf{P}_i\, \nabla \mathbf{u}\, \mathbf{P}_i\, .
\end{align}
Except notation, Eq.~\eqref{eqn:sameashulsen} is the same formulation as given in~\cite[Eq.~(44)]{Hulsen2005}.

To prove the equivalence to the most widespread log-conformation formulation, we introduce
\begin{align*}
	 \mathbf{\tilde{M}} = \begin{pmatrix} \tilde{m}_{11} & \tilde{m}_{12} & \tilde{m}_{13}\\
		\tilde{m}_{21} & \tilde{m}_{22} & \tilde{m}_{23}\\
	\tilde{m}_{31} & \tilde{m}_{32} & \tilde{m}_{33} \end{pmatrix}
		&\coloneqq \mathbf{O}^T \nabla \mathbf{u} \mathbf{O}\, .
\end{align*}
We can thus express $\mathbf{B}$ as
\begin{align}
    \mathbf{B} &= \mathbf{O} \begin{pmatrix} \tilde{m}_{11} & 0 & 0\\
			0 & \tilde{m}_{22} & 0\\
	0 & 0 & \tilde{m}_{33} \end{pmatrix} \mathbf{O}^T \, .
\end{align}
Moreover, considering the case of distinct eigenvalues, we introduce
\begin{align}
    \mathbf{\Omega} =& -\sum_{i\neq j} \frac{1}{e^{\lambda_i}-e^{\lambda_j}}
            \mathbf{P}_i \left(e^{\lambda_j} \nabla\mathbf{u}
                + e^{\lambda_i} \nabla \mathbf{u}^T \right)\mathbf{P}_j\, .
\end{align}
With the projection operators $\mathbf{P}_i$ being orthogonal and idempotent, i.e., $\mathbf{P}_i \mathbf{P}_j = \delta_{ij} \mathbf{P}_i$ and $\delta_{ij}$ being the Kronecker Delta, this yields the equivalent formulation
\begin{align}
\begin{split}
    &\mathbf{\Psi} \omega(\mathbf{u}) - \omega(\mathbf{u}) \mathbf{\Psi}
            - 2\, f(\opad \mathbf{\Psi})\, \epsilon(\mathbf{u})\\
        &\qquad= -2 \mathbf{B} + \mathbf{\Psi} \mathbf{\Omega} - \mathbf{\Omega} \mathbf{\Psi}\, .
\end{split}
\end{align}
Similarly to the formulation of $\mathbf{B}$ we can also reformulate $\mathbf{\Omega}$ using $\mathbf{\tilde{M}}$ as
\begin{align}
    \mathbf{\Omega} =& \mathbf{O} \begin{pmatrix} 0 & \omega_{12} & \omega_{13} \\
				\omega_{21} & 0 & \omega_{23} \\
            \omega_{31} & \omega_{32} & 0 \end{pmatrix} \mathbf{O}^T\, ,
\end{align}
where the $\omega_{ij}$ are given by
\begin{align}
	\omega_{ij} &\coloneqq - \frac{e^{\lambda_j} \tilde{m}_{ij} + e^{\lambda_i} \tilde{m}_{ji}}
			{e^{\lambda_i}-e^{\lambda_j}}\, .
\end{align}
This is mostly the original formulation, as it was first used by Fattal and Kupferman~\cite{Fattal2004} and has been used in many numerical implementations.

For the sake of completeness, and without proof, we also mention the formulation using a Dunford-type/Cauchy-type integral
\begin{align}
\begin{split}
    &f(\opad \mathbf{\Psi})\, \epsilon(\mathbf{u}) = \frac{1}{(2\pi i)^2} \times\\
        &\int_{\Gamma} \int_{\Gamma} f(z-z') \left(z\mathbf{1} - \mathbf{\Psi}\right)^{-1}
            \epsilon(\mathbf{u}) \left(z'\mathbf{1} - \mathbf{\Psi}\right)^{-1}\, dz\, dz'\, ,
\end{split}
\end{align}
where $\Gamma$ is a suitably chosen integration contour in the complex plane that encompasses the eigenvalues $\lambda_i$, but avoids the singularities of $f$. This formulation, which was proven in~\cite{Knechtges2015,Knechtges2018}, facilitates analytical insights into the log-conformation formulation, but is less suited for the direct numerical implementation, due to the expected cancellation effects in the Cauchy-type integral.

\section{Eigenvalue-Free Algorithm Design}\label{sec:eigVal-Free_Algo}
In the last section, we discussed several of the different existing formulations for the $f(\opad \mathbf{\Psi})\, \epsilon(\mathbf{u})$ term in the logarithmic constitutive equation. We also mentioned the connection to the eigenvalue-based algorithms.

In this section, we will come to an eigenvalue-free algorithm that represents $\opad \mathbf{\Psi}$ as a matrix on a suitably chosen vector space, which allows us to evaluate $f(\opad\mathbf{\Psi})$ as a matrix function.

Since it is instructive for what comes, and since it is also necessary for the numerical implementation, we will first concern ourselves with the eigenvalue-free evaluation of the matrix function $\exp(\mathbf{\Psi})$. We will use the Scaling\&Squaring algorithm, which has been extensively studied. For an in-depth review article, we refer to~\cite{Moler2003}.

The basic idea of the Scaling\&Squaring algorithm consists of two ingredients: One ingredient is a simple approximation of the function. This can, e.g., be a truncated Taylor series, or a rationale approximation. For the exponential function, the Padé approximation $R_{m,m}$ as a rationale approximation is a common choice. Usually, such an approximation is only reasonable in a small region close to some pivot point, which, for our approximation of the exponential function, is the origin of the coordinate system.

At this point, the second ingredient comes into action: a functional relation that helps to map the argument to the region where the aforementioned simple approximation is valid, and thus allows us to construct a more universal approximation. For the exponential function this relation is
\begin{align}
    \exp(\mathbf{\Psi}) =& \exp(\mathbf{\Psi}/2)^2\, .
\end{align}
Given a general $\mathbf{\Psi}$ and iterating this functional equation, one can choose a $j\in\mathbb{N}$ such that $2^{-j} \norm{\mathbf{\Psi}}$ is small enough for the Padé approximant to be sufficiently good. Then evaluating
\begin{align}
    \exp(\mathbf{\Psi}) \approx& \left(R_{m,m}(\mathbf{\Psi}/2^j)\right)^{2^j}
\end{align}
should give a reasonable approximation even for large $\mathbf{\Psi}$.
As is apparent, we first scale the argument with $2^{-j}$ and after the evaluation of the Padé approximant, we employ $j$ successive squarings. Hence, the name of the algorithm: Scaling\&Squaring.

For the actual implementation, we use the software library Eigen~\cite{eigenv3}, which uses variations of the Scaling\&Squaring algorithm, as described in~\cite[Algorithm~2.3]{Higham2005} and~\cite[Algorithm~3.1]{Almohy2010}.

In principle, as already noted in the previous section, we want to employ a similar algorithm for $f(\opad \mathbf{\Psi})$. However, we want to reduce the computational complexity first, i.e., we do not want to represent $\opad \mathbf{\Psi}$ as a $9\times 9$-matrix.

Notice that we will apply $f(\opad \mathbf{\Psi})$ to the symmetric matrix $\epsilon(\mathbf{u})$ and will get as a result a symmetric matrix again. In fact, the application of a symmetric matrix to $f(\opad \mathbf{\Psi})$ will always give a symmetric matrix. This can, e.g., be seen from Eq.~\eqref{eqn:firsteigenvallogconf} by simply transposing the equation, but also from the Taylor series expansion in Eq.~\eqref{eqn:taylorfad} and the fact that $\opad^2 \mathbf{\Psi}$ also has this feature: $\opad^2 \mathbf{\Psi}$ maps symmetric matrices to symmetric matrices, and antisymmetric matrices to antisymmetric matrices.

The fact that $\opad^2 \mathbf{\Psi}$ decomposes into two parts, of course nurtures the idea of just using the part operating on symmetric matrices. Since the vector space of symmetric $3\times 3$ matrices is only $6$-dimensional, this would already reduce the computational complexity. We could represent $\opad^2 \mathbf{\Psi}$ as a $6\times 6$-dimensional matrix, and then apply the function
\begin{align}\label{eqn:defg}
    g(z) =& \begin{cases} \frac{\sqrt{z}}{\tanh\sqrt{z}} & \mbox{for } \opre{z} \geq 0\\
            \frac{\sqrt{-z}}{\tan\sqrt{-z}} & \mbox{for } \opre{z} < 0\\
    \end{cases}\, ,
\end{align}
such that
\begin{align}
    f(\opad \mathbf{\Psi}) =& g\left(\frac{1}{4}\opad^2 \mathbf{\Psi} \right)\, .
\end{align}
This shifts the problem from evaluating a matrix function $f$ to a matrix function $g$.
Note that we have added the negative real part in Eq.~\eqref{eqn:defg} to illustrate that $g$ can be continued analytically in the negative half-plane to a meromorphic function. It thus becomes evident that $g$ has a pole at $z=-\pi^2$. In fact, the Taylor expansion follows from Eq.~\eqref{eqn:taylorf}
\begin{gather}
    g(z) = \sum_{n=0}^\infty \frac{B_{2n}}{(2n)!} 4^n z^n\, ,
\end{gather}
which, due to the pole, only converges absolutely for $\seminorm{z} < \pi^2$.

However, we can go one step further: First, we notice that $\opad \mathbf{\Psi}$ maps symmetric matrices, like $\epsilon(\mathbf{u})$, to an antisymmetric $3\times 3$-matrix. More importantly, the vector space of antisymmetric $3\times 3$-matrices is $3$-dimensional, hence any $\opad^{2n} \mathbf{\Psi}$ is at most of rank-$3$ as a linear operator or matrix for $n > 0$. In other words, in the Taylor series of $f$ or $g$ applied to $\opad \mathbf{\Psi}$, only the $n=0$ term, which is the identity operator/matrix $\mathbf{1}$, is of full rank, while all other terms are at most of rank-$3$.

This clearly motivates to split off the identity matrix~$\mathbf{1}$ and only compute the remaining part on a $3\times 3$-matrix instead of a $6\times 6$-matrix or $9\times 9$-matrix.
Thus, we define
\begin{gather}
    h(x) = \frac{1}{x} \left(\frac{\sqrt{x}}{\tanh \sqrt{x}} - 1\right)\, ,
\end{gather}
with its Taylor series for small $x$ given as
\begin{align}\label{eqn:taylorh}
    h(x) =& \sum_{n=1} \frac{B_{2n}}{(2n)!} 4^n x^{n-1}\, .
\end{align}
Using the equations above, we can write
\begin{align}
\label{eqn:fadpsi_without_eigenvaluedecomposition}
\begin{split}
    &f(\opad\mathbf{\Psi})\, \epsilon(\mathbf{u})\\
    &\qquad = \epsilon(\mathbf{u}) + \frac{1}{4} \opad\mathbf{\Psi}\,\, h\left(\frac{1}{4} \opad^2 \mathbf{\Psi}\right)\,\, \opad \mathbf{\Psi}\, \epsilon(\mathbf{u})\, ,
\end{split}
\end{align}
which contains already all components of the final algorithm that will compute $f(\opad \mathbf{\Psi}) \epsilon(\mathbf{u})$.

In the actual computation, we will need different representations of $\opad \mathbf{\Psi}$. Going through the different instances of $\opad \mathbf{\Psi}$ in Eq.~\eqref{eqn:fadpsi_without_eigenvaluedecomposition} from right to left:
\begin{itemize}
\item $\opad \mathbf{\Psi} \epsilon(\mathbf{u})$ as noted earlier is an antisymmetric $3\times 3$-matrix, and thus can be represented in some basis as a $3$-dimensional vector. We will denote this vector as $\mathbf{v}\in\mathbb{R}^3$.
\item On the $3$-dimensional space of antisymmetric $3\times 3$-matrices, the operator $\opad^2\mathbf{\Psi}$ will be represented as a $3\times 3$-matrix, which we will denote by $\mathbf{X}\in\mathbb{R}^{3\times 3}$. Dividing by four and applying $h$ gives another $3\times 3$-matrix $h\left(\frac{1}{4}\opad^2 \mathbf{\Psi}\right)$, which is multiplied with the $3$-vector $\mathbf{v}$ that represents $\opad \mathbf{\Psi} \epsilon(\mathbf{u})$. The final result of $h\left(\frac{1}{4}\opad^2 \mathbf{\Psi}\right) \opad \mathbf{\Psi} \epsilon(\mathbf{u})$ is once again then represented by a $3$-vector $h(\mathbf{X}/4)\, \mathbf{v}$.
\item The last invocation of $\opad \mathbf{\Psi}$ linearly maps an antisymmetric $3\times 3$-matrix to a symmetric $3\times 3$-matrix. Therefore, it can be represented as a $6\times 3$-matrix, which we will denote by $\mathbf{Y}\in\mathbb{R}^{6\times 3}$. It is multiplied by the $3$-vector from the previous step, resulting in a 6-vector $\mathbf{Y}\,h(\mathbf{X}/4)\, \mathbf{v}$.
\end{itemize}

In order to concretize the computational steps, we will need to choose specific bases. We start with the basis for the symmetric $3\times 3$-matrices: the matrix $\mathbf{\Psi}$ is already stored in most codes as a $6$-vector $(\Psi_{11},\Psi_{12},\Psi_{13},\Psi_{22},\Psi_{23},\Psi_{33})^T$. The same holds for $\epsilon(\mathbf{u})$ with $(\epsilon_{11},\epsilon_{12},\epsilon_{13},\epsilon_{22},\epsilon_{23},\epsilon_{33})^T$.

For the antisymmetric $3\times 3$-matrices to be represented as a $3$-vector, we want to have further properties for the representation of $\opad^2 \mathbf{\Psi}$ as a matrix on that vector space. Most notably, we want $\opad^2 \mathbf{\Psi}$ to be represented as a symmetric matrix $\mathbf{X}\in\mathbb{R}^{3\times 3}_{sym}$.

For that, we first define a scaled Frobenius product of two matrices $\mathbf{A},\mathbf{B}\in \mathbb{R}^{3\times 3}$, i.e.,
\begin{gather}
    \sprod{\mathbf{A}}{\mathbf{B}}_{sF} \coloneqq \frac{1}{2} \optr{\mathbf{A}^T \mathbf{B}}\, .
\end{gather}
The factor $1/2$ is introduced to avoid several $\sqrt{2}$ factors in the following formulas. The more important aspect here is that $\opad^2 \mathbf{\Psi}$ is selfadjoint with respect to this scalar product
\begin{align*}
    \sprod{\mathbf{A}}{\opad^2 \mathbf{\Psi}\, \mathbf{B}}_{sF}
        =& \frac{1}{2} \optr \left(\mathbf{A}^T \left(\mathbf{\Psi}^2 \mathbf{B} - 2 \mathbf{\Psi} \mathbf{B} \mathbf{\Psi} + \mathbf{B} \mathbf{\Psi}^2\right)\right)\\
        =& \frac{1}{2} \optr \left(\left(\mathbf{\Psi}^2 \mathbf{A} - 2 \mathbf{\Psi} \mathbf{A} \mathbf{\Psi} + \mathbf{A} \mathbf{\Psi}^2\right)^T \mathbf{B} \right)\\
        =& \sprod{\opad^2 \mathbf{\Psi}\, \mathbf{A}}{\mathbf{B}}_{sF}\, .
\end{align*}
One ramification of the selfadjointness is that the matrix representation of $\opad^2 \mathbf{\Psi}$ in a basis will yield a symmetric matrix $\mathbf{X}$, if we choose that basis to be orthonormal with respect to the same scalar product.

This motivates our choice of an orthonormal basis $\left\{\mathbf{E}_i\right\}$ of the antisymmetric $3\times 3$-matrices
\begin{align}
    \mathbf{E}_1 =& \begin{pmatrix} 0 & 1 & 0 \\ -1 & 0 & 0 \\ 0 & 0 & 0 \end{pmatrix}\\
    \mathbf{E}_2 =& \begin{pmatrix} 0 & 0 & 1 \\ 0 & 0 & 0 \\ -1 & 0 & 0 \end{pmatrix}\\
    \mathbf{E}_3 =& \begin{pmatrix} 0 & 0 & 0 \\ 0 & 0 & 1 \\ 0 & -1 & 0 \end{pmatrix}\, .
\end{align}

\begin{algorithm}[t]
\caption{Computing $f(\opad\mathbf{\Psi})\, \epsilon(\mathbf{u})$}
\label{alg:fopad}
\begin{algorithmic}
\REQUIRE $\mathbf{\Psi}$ given as $(\Psi_{11},\Psi_{12},\Psi_{13},\Psi_{22},\Psi_{23},\Psi_{33})^T$
\REQUIRE $\epsilon(\mathbf{u})$ given as $(\epsilon_{11},\epsilon_{12},\epsilon_{13},\epsilon_{22},\epsilon_{23},\epsilon_{33})^T$
\STATE compute $\mathbf{v}\in\mathbb{R}^3$ according to Eq.~\eqref{eqn:defv1}--\eqref{eqn:defv3}
\STATE compute $\mathbf{X}\in\mathbb{R}^{3\times 3}$ according to Eq.~\eqref{eqn:defX11}--\eqref{eqn:defX33}
\STATE compute $\mathbf{Y}\in\mathbb{R}^{6\times 3}$ according to Eq.~\eqref{eqn:defY}
\STATE use Algorithm~\ref{alg:hx} to compute $\mathbf{Z} \gets h(\mathbf{X}/4)$
\RETURN $\epsilon(\mathbf{u}) + \frac{1}{4}\mathbf{Y}\, \mathbf{Z}\, \mathbf{v}$
\end{algorithmic}
\end{algorithm}

Going through the different needed representations of $\opad\mathbf{\Psi}$, we will start with $\opad \mathbf{\Psi}\, \epsilon(\mathbf{u})$, which we represent as a vector $\mathbf{v}\in \mathbb{R}^3$, whose components are given by $v_i = \sprod{\mathbf{E}_i}{\opad \mathbf{\Psi}\, \epsilon(\mathbf{u})}_{sF}$. The latter yields
\begin{align}\label{eqn:defv1}
\begin{split}
    v_1 =& -\epsilon_{11} \Psi_{12} + \epsilon_{12} \Psi_{11} - \epsilon_{12}\Psi_{22}\\
        &\qquad- \epsilon_{13}\Psi_{23} + \epsilon_{22}\Psi_{12} + \epsilon_{23}\Psi_{13}
\end{split}\\
\begin{split}
    v_2 =& -\epsilon_{11}\Psi_{13} - \epsilon_{12}\Psi_{23} + \epsilon_{13}\Psi_{11}\\
        &\qquad - \epsilon_{13}\Psi_{33} + \epsilon_{23}\Psi_{12} + \epsilon_{33}\Psi_{13}
\end{split}\\
\label{eqn:defv3}
\begin{split}
    v_3 =& -\epsilon_{12}\Psi_{13} + \epsilon_{13}\Psi_{12} - \epsilon_{22}\Psi_{23}\\
        &\qquad + \epsilon_{23}\Psi_{22} - \epsilon_{23}\Psi_{33} + \epsilon_{33}\Psi_{23}\, .
\end{split}
\end{align}

To represent $\opad^2\mathbf{\Psi}$ on the space of antisymmetric $3\times 3$-matrices, we introduce $\mathbf{X} \in \mathbb{R}^{3\times 3}$, whose entries are given by $X_{ij} = \sprod{\mathbf{E}_i}{\opad^2\mathbf{\Psi}\, \mathbf{E}_j}_{sF}$. As noted, the resulting matrix $\mathbf{X}$ is symmetric. With our chosen basis, the coefficients are given by
\begin{align}\label{eqn:defX11}
    X_{11} =& \Psi_{11}^2 - 2 \Psi_{11}\Psi_{22} + 4 \Psi_{12}^2 + \Psi_{13}^2 + \Psi_{22}^2 + \Psi_{23}^2\\
    X_{12} =& -2\Psi_{11}\Psi_{23} + 3\Psi_{12}\Psi_{13} + \Psi_{22}\Psi_{23} + \Psi_{23}\Psi_{33}\\
    X_{13} =& -\Psi_{11}\Psi_{13} - 3\Psi_{12}\Psi_{23} + 2\Psi_{13}\Psi_{22} - \Psi_{13}\Psi_{33}\\
    X_{22} =& \Psi_{11}^2 - 2\Psi_{11}\Psi_{33} + \Psi_{12}^2 + 4\Psi_{13}^2 + \Psi_{23}^2 + \Psi_{33}^2\\
    X_{23} =& \Psi_{11}\Psi_{12} + \Psi_{12}\Psi_{22} - 2\Psi_{12}\Psi_{33} + 3\Psi_{13}\Psi_{23}\\
    \label{eqn:defX33}
    X_{33} =& \Psi_{12}^2 + \Psi_{13}^2 + \Psi_{22}^2 - 2\Psi_{22}\Psi_{33} + 4\Psi_{23}^2 + \Psi_{33}^2\, .
\end{align}

For the last representation of $\opad\mathbf{\Psi}$, from the space of antisymmetric matrices to the space of symmetric $3\times 3$-matrices, we compute $\opad\mathbf{\Psi}\, \mathbf{E}_i$ and extract the coefficients. We denote the representation by $\mathbf{Y}\in \mathbb{R}^{6\times 3}$ and its coefficients are given by
\begin{align}\label{eqn:defY}
    \mathbf{Y} =& \begin{pmatrix}
        -2\Psi_{12} & -2\Psi_{13} & 0 \\
        \Psi_{11}-\Psi_{22} & - \Psi_{23} & -  \Psi_{13}\\
        - \Psi_{23} & \Psi_{11}-\Psi_{33} & \Psi_{12}\\
        2\Psi_{12} & 0 & -2\Psi_{23} \\
        \Psi_{13} & \Psi_{12} & \Psi_{22}-\Psi_{33}\\
        0 &2\Psi_{13} & 2\Psi_{23}
    \end{pmatrix}\, .
\end{align}

Taking for the moment the algorithm to compute $h(\mathbf{X}/4)$ as given, we can then use Eq.~\eqref{eqn:fadpsi_without_eigenvaluedecomposition} to compute $f(\opad\mathbf{\Psi})\,\epsilon(\mathbf{u})$ as a series of matrix operations. The actual algorithm to compute $f(\opad\mathbf{\Psi})\epsilon(\mathbf{u})$ is illustrated in Algorithm~\ref{alg:fopad}.

Before we come to the general case of computing $h(\mathbf{X}/4)$, we want to first mention a case in which evaluating $h$ becomes as easy as a simple function evaluation: the two-dimensional case.

To see this, note that in the two-dimensional case $X_{12}$ and $X_{13}$ are both zero. As such $\mathbf{X}$ is the direct sum of two submatrices, of which the first one consists of just a single entry $X_{11}$. Furthermore, since $v_1$ is the only non-zero entry of $\mathbf{v}$ in this case, it is also just $h(X_{11})$ that needs to be calculated. In fact, acknowledging that
\begin{align}
    h(x) =& \frac{1}{x}\left(\sqrt{x} + \frac{2\sqrt{x}}{e^{2\sqrt{x}}-1}-1\right)\, ,
\end{align}
this yields exactly the representation of $f(\opad\mathbf{\Psi})\,\epsilon(\mathbf{u})$ that was given in~\cite[Theorem~2]{Knechtges2014}.

\begin{algorithm}[t]
\caption{Computing $h(\mathbf{X})$}
\label{alg:hx}
\begin{algorithmic}
\REQUIRE $\mathbf{X}$
\STATE $j\gets \max \left(0, j_0 + \mbox{\texttt{std::ilogb}} (\norm{\mathbf{X}}_F^2)/4\right)$
\STATE $\mathbf{L} \gets \mathbf{X}/4^{j}$
\STATE $\mathbf{H} \gets \left(\left(\left(\frac{5}{66} \frac{4^5}{3628800} \mathbf{L} - \frac{1}{30} \frac{4^4}{40320} \mathbf{1}\right) \mathbf{L} + \frac{1}{42}\frac{4^3}{720} \mathbf{1}\right) \mathbf{L} \right.$\\$\qquad\qquad\left.- \frac{1}{30} \frac{4^2}{24} \mathbf{1}\right) \mathbf{L} + \frac{1}{6} \frac{4}{2} \mathbf{1}$
\STATE $\mathbf{G} \gets \mathbf{1} + \mathbf{H}\, \mathbf{L}$
\FOR{$i=1$ \TO $j$}
\STATE $\mathbf{H} \gets \frac{1}{4} \left(\mathbf{H} + \mathbf{G}^{-1}\right)$
\STATE $\mathbf{L} \gets 4\, \mathbf{L}$
\STATE $\mathbf{G} \gets \mathbf{1} + \mathbf{H}\, \mathbf{L}$
\ENDFOR
\RETURN $\mathbf{H}$
\end{algorithmic}
\end{algorithm}

Coming to the general case, we so far only have a Taylor series of $h$, Eq.~\eqref{eqn:taylorh}, which only works for small $\mathbf{X}$. Taking the Scaling\&Squaring algorithm for the matrix exponential function as an instructive example, we seek a functional equation that allows us to reduce the computation to arguments that are amenable to the Taylor series.

In fact, using the formula for doubling the argument of $\tanh x$
\begin{align}
    \tanh x =& \frac{2\tanh \frac{x}{2}}{1+\tanh^2 \frac{x}{2}}\, ,
\end{align}
we obtain
\begin{align}\label{eqn:iteratingh}
    h(x) =& \frac{1}{4}\left(h(x/4) + \left(g(x/4)\right)^{-1}\right)\,
\end{align}
with
\begin{align}\label{eqn:iteratingg}
    g(x/4) = 1+x/4\, h(x/4)\, .
\end{align}
Considering a general $\mathbf{X}\in \mathbb{R}^{3\times 3}_{sym}$, we can readily use this to seek an appropriate $j\in\mathbb{N}$ such that $\mathbf{X}/4^j$ is small enough to be approximated by a truncated Taylor series. Then iterating Eqs.~\eqref{eqn:iteratingh} and \eqref{eqn:iteratingg} $j$-times we get the final result. The full algorithm is displayed in Algorithm~\ref{alg:hx}.

For the actual algorithm, we needed to decide on when $\mathbf{X}/4^j$ is small enough. Evaluating Algorithm~\ref{alg:hx} for scalar instead of matrix arguments and comparing it with a high-precision calculation of $h$ gives an indication on the accuracy of the algorithm. This analysis yields that $j_0=4$ is sufficient for an absolute accuracy of $10^{-16}$ in the scalar argument case. Therefore, this is also the value that was used for all our numerical evaluations. We also compared the matrix argument case with an eigenvalue-based evaluation for random $\mathbf{X}$ and could not observe any severe issues.

However, this should not be taken without a word of caution. The functions $g(x)$ and $h(x)$ asymptotically behave like $\sqrt{x}$ and $1/\sqrt{x}$, respectively. In fact, $\sqrt{x}$ as a function is known as a prime example, where an algorithm works for scalar arguments, but may fail for matrix arguments of even moderate condition number, cf.~\cite{higham1986newton,Higham2008}. Hence, although our numerical experiments do already give a strong indication for a stable algorithm, a thorough mathematical error analysis of the algorithm is still outstanding and subject of future research.

\section{Finite Volume Implementation}\label{sec:numimpl}
In the following, we are going to embed the new eigenvalue-free constitutive formulation in a numerical implementation. We will, therefore, augment the constitutive equation~\eqref{eqn:logconf} with a system of partial differential equations consisting of the continuity equation and the momentum balance, as well as Kramers' expression to relate the polymeric stress and the log-conformation field. These equations will then be solved using a finite volume method (FVM), where the polymeric stress is computed with the log-conformation approach according to Eq.~\eqref{eqn:logconf}, and where the $f(\opad \mathbf{\Psi})\, \epsilon(\mathbf{u})$-term on the right-hand side is computed without an eigenvalue decomposition of $\mathbf{\Psi}$ according to Eq.~\eqref{eqn:fadpsi_without_eigenvaluedecomposition} and Algorithm~\ref{alg:fopad}.

As noted earlier, the eigenvalue-free log-conformation formulation is quite universal and not necessarily tied to a specific discretization scheme. Like the eigenvalue-based formulation, it needs a point-based evaluation of $\mathbf{\Psi}$, and the discretization scheme needs to provide a good approximation of $\nabla \mathbf{u}$ at the same point, such that Algorithm~\ref{alg:fopad} can compute the $f(\opad \mathbf{\Psi})\, \epsilon(\mathbf{u})$ term also at this point in space and time.

To illustrate how easily this different evaluation of the $f(\opad \mathbf{\Psi})\, \epsilon(\mathbf{u})$ term can be dropped into an existing code, we chose to base our numerical implementation on one of the existing and established open source computational rheology packages: RheoTool~\cite{rheoTool}. It is based on OpenFOAM\textsuperscript{\textregistered}~\cite{Weller1998} and has many constitutive models for viscoelastic fluid simulations implemented already. The eigenvalue-free formulations for the log-conf variants of the Oldroyd-B and Giesekus models, which are subject of this work, are implemented among those models and can be used and configured analogously in the overall OpenFOAM\textsuperscript{\textregistered} framework. More specifically, we use RheoTool in version 6 and OpenFOAM\textsuperscript{\textregistered} in version 9.

A detailed description of the system of partial differential equations and algebraic equations that we will use, and of the corresponding finite volume discretization and linearization follows next. Afterwards, in Section~\ref{sec:benchmarks}, our implementation is applied to the study of two well-known tests for viscoelastic fluid flow: the confined cylinder and the sedimenting sphere benchmarks.

\subsection{Statement of the full set of partial differential equations}
To state the full system of partial differential equations, which we are going to discretize and solve, we start with the incompressible isothermal Navier--Stokes equations
\begin{gather}
    \nabla \cdot \mathbf{u} = 0 \label{eqn:continuity} \displaybreak[0] \\
    \rho (\partial_t\mathbf{u} + (\mathbf{u} \cdot \nabla) \mathbf{u}) = -\nabla p + \nabla \cdot (\eta_s \nabla \mathbf{u}) + \nabla \cdot \bm{\tau}\, , \label{eqn:momentum}
\end{gather}
where $\mathbf{u}$ is the velocity vector, $p$ the pressure, $\bm{\tau}$ the polymeric extra stress tensor, $\eta_s$ the solvent viscosity, and $\rho$ the density of the fluid. These equations are coupled with an additional partial differential equation for the log-conf tensor $\mathbf{\Psi}$, which was already stated in its $f(\opad \mathbf{\Psi})$ form in Eq.~\eqref{eqn:logconf}. Rearranging some terms, the constitutive equation can be written as
\begin{gather}\label{eqn:logconf_sec4}
\begin{split}
	\partial_t \mathbf{\Psi} + (\mathbf{u} \cdot \nabla) \mathbf{\Psi}
		= -\mathbf{\Psi} \omega(\mathbf{u}) + \omega(\mathbf{u}) \mathbf{\Psi}\\\qquad
		+ 2\, f(\opad \mathbf{\Psi})\, \epsilon(\mathbf{u})
		- P(e^{\mathbf{\Psi}}) e^{-\mathbf{\Psi}}\, .
\end{split}
\end{gather}
In the following benchmarks, we only consider the Oldroyd-B and Giesekus constitutive models, thus setting $P(\exp(\mathbf{\Psi})) = \frac{1}{\lambda} \left(\exp(\mathbf{\Psi}) - \mathbf{1}\right)$ and $P(\exp(\mathbf{\Psi})) = \frac{1}{\lambda} \left(\mathbf{1} + \alpha \left(\exp(\mathbf{\Psi}) - \mathbf{1}\right)\right) \left(\exp(\mathbf{\Psi}) - \mathbf{1}\right)$, respectively.
The conformation tensor and the log-conformation field are related to the polymeric stress $\bm{\tau}$ by means of Kramers' expression
\begin{equation}\label{eqn:psitotau}
    \bm{\tau} = \frac{\eta_p}{\lambda} (e^\mathbf{\Psi} - \mathbf{1})\, ,
\end{equation}
where $\eta_p$ is the polymeric viscosity and $\lambda$ the relaxation time of the fluid.

In total, the set of partial differential equations and one algebraic equation~\eqref{eqn:continuity}--\eqref{eqn:psitotau} composes, when augmented with appropriate initial and boundary conditions, the mathematical problem we try to solve. In the following, we will lay out our chosen discretization scheme.

\subsection{Temporal discretization, linearization and SIMPLEC}
Starting off with the set of equations in Eqs.~\eqref{eqn:continuity}--\eqref{eqn:psitotau}, we at first discretize in time using the backwards Euler scheme. This leads to
\begin{gather}
    \nabla \cdot \mathbf{u}_t = 0 \displaybreak[0] \\
\begin{split}
    \frac{\rho}{\Delta t} \mathbf{u}_t + \rho \left(\mathbf{u}_t\cdot \nabla\right) \mathbf{u}_t
        &= -\nabla p_t + \nabla\cdot(\eta_s\nabla \mathbf{u}_t)\\&\qquad
        + \nabla \cdot \bm{\tau}_t
        + \frac{\rho}{\Delta t} \mathbf{u}_{t-\Delta t}
\end{split} \displaybreak[0] \\
\begin{split}
    \frac{1}{\Delta t} \mathbf{\Psi}_t + (\mathbf{u}_t \cdot \nabla) \mathbf{\Psi}_t
	&= -\mathbf{\Psi}_t \omega(\mathbf{u}_t) + \omega(\mathbf{u}_t) \mathbf{\Psi}_t\\&\qquad
	+ 2\, f(\opad \mathbf{\Psi}_t)\, \epsilon(\mathbf{u}_t)\\&\qquad
	- P(e^{\mathbf{\Psi}_t}) e^{-\mathbf{\Psi}_t}
        + \frac{1}{\Delta t} \mathbf{\Psi}_{t-\Delta t}
\end{split} \displaybreak[0] \\
    \bm{\tau}_t = \frac{\eta_p}{\lambda} (e^{\mathbf{\Psi}_t} - \mathbf{1})\, . \label{eqn:psitotau_t}
\end{gather}
In order to not overload the notation, we drop the $t$ indices from the current time-step and only keep $\mathbf{u}_{t-\Delta t}$ and $\mathbf{\Psi}_{t-\Delta t}$.

As a next step, we approach the non-linearity. Therefore, we choose a Picard-type fixed-point iteration. We indicate the current iteration with a suffix $i$ and start our iteration with $\mathbf{u}_0 = \mathbf{u}_{t-\Delta t}$ and $\mathbf{\Psi}_0 = \mathbf{\Psi}_{t-\Delta t}$. We linearize our equations in such a way that $\mathbf{\Psi}_i$ is solved for after $\mathbf{u}_i$ and $p_i$ have been computed. In the constitutive equation, all non-linear occurrences of $\mathbf{\Psi}$ are replaced by $\mathbf{\Psi}_{i-1}$. In the momentum equation, we choose to linearize the convective derivative as usual, by computing the flux based on the previous iteration.
We thus obtain
\begin{gather} \label{eqn:continuity_i}
    \nabla \cdot \mathbf{u}_i = 0 \displaybreak[0] \\
\begin{split}
    \frac{\rho}{\Delta t} \mathbf{u}_i + \rho \left(\mathbf{u}_{i-1}\cdot \nabla\right) \mathbf{u}_i
        &= -\nabla p_i + \nabla\cdot(\eta_s\nabla \mathbf{u}_i)\\&\qquad
            + \nabla \cdot \bm{\tau}_{i-1}
            + \frac{\rho}{\Delta t} \mathbf{u}_{t-\Delta t} \label{eqn:momentum_i}
\end{split} \displaybreak[0] \\
\begin{split}
	\frac{1}{\Delta t} \mathbf{\Psi}_i + (\mathbf{u}_i \cdot \nabla) \mathbf{\Psi}_i
		&= -\mathbf{\Psi}_{i-1} \omega(\mathbf{u}_i) + \omega(\mathbf{u}_i) \mathbf{\Psi}_{i-1}\\&\qquad
		+ 2\, f(\opad \mathbf{\Psi}_{i-1})\, \epsilon(\mathbf{u}_i)\\&\qquad
		- P(e^{\mathbf{\Psi}_{i-1}}) e^{-\mathbf{\Psi}_{i-1}}
     + \frac{1}{\Delta t} \mathbf{\Psi}_{t-\Delta t} \label{eqn:log-conf_i}
\end{split} \displaybreak[0] \\
    \bm{\tau}_i = \frac{\eta_p}{\lambda} (e^{\mathbf{\Psi}_i} - \mathbf{1})\, . \label{eqn:psi-to-tau_i}
\end{gather}

Note that the way we have linearized the system, first $\mathbf{u}_i$ and $p_i$ should be solved in a coupled way, then $\mathbf{\Psi}_i$ can be computed based on $\mathbf{u}_i$, which, at last, results in $\bm{\tau}_i$.
It is also noteworthy that our chosen scheme does not use any type of both-sides diffusion (BSD), which was introduced in \cite{guenette1995new} and applied in a finite volume context in \cite{pimenta2017stabilization}.

To further reduce the coupling between $\mathbf{u}_i$ and $p_i$ we employ the SIMPLEC method~\cite{vandoormaal1984enhancements}. For that, consider the following form of the momentum equation~\eqref{eqn:momentum_i}
\begin{align}\label{eqn:momentum_ah}
\begin{split}
    &\overbrace{\left(\frac{\rho}{\Delta t} + \rho (\mathbf{u}_{i-1}\cdot \nabla) - \nabla \cdot(\eta_s \nabla) \right)}^{\eqqcolon A - H} \mathbf{u}^* \\&\qquad\qquad\qquad= - \nabla p^*
        + \underbrace{\nabla \cdot \bm{\tau}_{i-1} + \frac{\rho}{\Delta t} \mathbf{u}_{t-\Delta t}}_{\eqqcolon \mathbf{b}}\, ,
\end{split}
\end{align}
where $A - H$ encodes the linear operator that operates on $\mathbf{u}$ in the momentum equation.\footnote{Our notation deviates a bit from the actual implementation in OpenFOAM\textsuperscript{\textregistered}, where $H$ is used to denote what is here given as $H\mathbf{u}^* + \mathbf{b}$.} After the spatial discretization, which follows in Section~\ref{sec:FVMdiscretization}, $A$ will be the diagonal part of the matrix and $-H$ the off-diagonal part. In particular $A$ will be easy to invert.

Now, assuming $\mathbf{u}^*$ solves Eq.~\eqref{eqn:momentum_ah} given the pressure $p^* \coloneqq p_{i-1}$ from the previous iteration, we seek an update $\mathbf{u}'$ such that $\mathbf{u}_i = \mathbf{u}^* + \mathbf{u}'$ solves the continuity equation~\eqref{eqn:continuity_i}. Introducing the pressure update $p' = p_i - p^*$, the velocity update $\mathbf{u}'$ needs to solve
\begin{align}
    \left(A - H\right) \mathbf{u}' &= - \nabla p'\, .
\end{align}
SIMPLEC now approximates $H$ by another operator $H_1$, which like $A$ is easy to invert. In the actual implementation, i.e., after the spatial discretization, $H_1$ will be realized as a matrix lumping of the off-diagonal entries onto the diagonal. For the details consult~\cite{pimenta2017stabilization}. Thus, we can solve
\begin{align}
    \mathbf{u}' &= - \left(A-H_1\right)^{-1} \nabla p' \, .
\end{align}
Therefore, the continuity equation $\nabla \cdot (\mathbf{u}' + \mathbf{u}^*) = 0$ amounts to
\begin{align}\label{eqn:pre_pressure_correction}
\begin{split}
    0 = \nabla \cdot &\left(- \left(A-H_1\right)^{-1} \nabla (p_i - p^*)\right.\\&\quad
        \left.+ A^{-1} \left(H\mathbf{u}^* - \nabla p^* + \mathbf{b}\right)\right)\, ,
\end{split}
\end{align}
which can be rearranged to the pressure correction equation
\begin{align}\label{eqn:pressure_correction}
\begin{split}
    &\nabla \cdot \left(\left(A-H_1\right)^{-1} \nabla p_i\right)\\
        &\quad= \nabla \cdot \left(A^{-1} (H\mathbf{u}^* + \mathbf{b})
            +\left((A-H_1)^{-1} - A^{-1}\right) \nabla p^* \right)\, .
\end{split}
\end{align}
The corrected velocity $\mathbf{u}_i$ is then given by
\begin{align}\label{eqn:velocity_correction}
\begin{split}
    \mathbf{u}_i &= A^{-1} (H\mathbf{u}^* + \mathbf{b})
            +\left((A-H_1)^{-1} - A^{-1}\right) \nabla p^*\\&\qquad
            - \left(A-H_1\right)^{-1} \nabla p_i\, .
\end{split}
\end{align}

\begin{figure*}[p!]
\centering
\footnotesize
\selectcolormodel{gray}

\scalebox{0.84}{
\begin{tikzpicture}[node distance=0.5cm]

\tikzstyle{title} = [font=\color{black}\scshape]

\tikzstyle{process} = [rectangle, minimum height=1cm, text centered, text width=7cm, draw=black, fill=white]
\tikzstyle{place} = [diamond, aspect=4, minimum size=2mm, thick, draw, fill=white]

\tikzstyle{pre}  = [<-,shorten >=1pt,>=stealth,semithick]
\tikzstyle{mid}  = [-,semithick]
\tikzstyle{post} = [->,>=stealth,semithick]

\node (init) {Initialize the fields $\{\mathbf{u}, p, \mathbf{\Psi}, \bm{\tau}\}_0$};

    \node (tl1) [below = of init] {$t\leftarrow0$} edge[mid] (init);
    \node (tl2) [below = of tl1, circle, draw, inner sep = 2pt] {} edge[pre] (tl1);

        \node (innerit1) [below = of tl2] {$i\leftarrow0$} edge[mid] (tl2);
        \node (innerit2aoihf) [below = of innerit1, circle, draw, inner sep = 2pt] {} edge[pre] (innerit1);
        \node (momentum) [process, below = of innerit2aoihf] {Solve Eq.~\eqref{eqn:momentum_ah} for the intermediate velocity $\mathbf{u}_i^*$} edge[pre] (innerit2aoihf);

                \node (innerit3) [below = of momentum] {$j\leftarrow0$} edge[mid] (momentum);
                \node (innerit4) [below = of innerit3, circle, draw, inner sep = 2pt] {} edge[pre] (innerit3);
                \node (continuity) [process, below = of innerit4] {Solve Eq.~\eqref{eqn:pressure_correction} for the continuity \\ compliant pressure $p_j$} edge[pre] (innerit4);
                \node[place] (finalNOLoop) [below = of continuity] {Final non-orthogonal correction?} edge[pre] (continuity);

                \node (jupdate) [right = 0.5cm of continuity] {$j\leftarrow j+1$};
                \draw [mid,draw] (finalNOLoop) -| (jupdate);
                \draw [post,draw] (jupdate) |- (innerit4);
                
                \coordinate (dummycoord1) at ($(finalNOLoop.east)$) edge[mid] (finalNOLoop);
                \draw [mid,draw] (finalNOLoop) -| (dummycoord1) node[near end, anchor= south west] {no};
                \node (dummynode1) [below = 1cm of finalNOLoop] {} edge[mid] (finalNOLoop);
                \coordinate (dummycoord2) at ($(finalNOLoop.south)$) edge[mid] (finalNOLoop);
                \draw [mid,draw] (dummycoord2) -| (dummynode1) node[near end, anchor= south west] {yes};

            \node (setpi) [below = 1cm of finalNOLoop] {$p_i\leftarrow p_j$} edge[mid] (finalNOLoop);
            \node (ucorrection) [process, below = of setpi] {Compute the corrected velocity $\mathbf{u}_i$ using Eq.~\eqref{eqn:velocity_correction}} edge[pre] (setpi);
            
        \node (logconf) [process, below = of ucorrection] {Solve Eq.~\eqref{eqn:log-conf_i} for the log-conf tensor $\mathbf{\Psi}_i$} edge[pre] (ucorrection);
        \node (psitotau) [process, below = of logconf] {Compute the polymeric stress $\bm{\tau}_i$ using Eq.~\eqref{eqn:psi-to-tau_i}} edge[pre] (logconf);
        \node[place] (finalinnerloop) [below = of psitotau] {Final inner iteration?} edge[pre] (psitotau);

        \node (iupdate) [right = 2.5cm of ucorrection] {$i\leftarrow i+1$};
        \draw [mid,draw] (finalinnerloop) -| (iupdate);
        \draw [post,draw] (iupdate) |- (innerit2aoihf);
        
        \coordinate (dummycoord3) at ($(finalinnerloop.east)$) edge[mid] (finalinnerloop);
        \draw [mid,draw] (finalinnerloop) -| (dummycoord3) node[near end, anchor= south west] {no};
        \node (dummynode2) [below = 1cm of finalinnerloop] {} edge[mid] (finalinnerloop);
        \coordinate (dummycoord4) at ($(finalinnerloop.south)$) edge[mid] (finalinnerloop);
        \draw [mid,draw] (dummycoord4) -| (dummynode2) node[near end, anchor= south west] {yes};
    
    \node (setfields) [below = 1cm of finalinnerloop] {$\{\mathbf{u}, p, \mathbf{\Psi}, \bm{\tau}\}_t \leftarrow \{\mathbf{u}, p, \mathbf{\Psi}, \bm{\tau}\}_i$} edge[mid] (finalinnerloop);
    \node[place] (finaltimestep) [below = of setfields] {Final timestep?} edge[pre] (setfields);

    \node (tupdate) [right = 4.5cm of logconf] {$t\leftarrow t+\Delta t$};
    \draw [mid,draw] (finaltimestep) -| (tupdate);
    \draw [post,draw] (tupdate) |- (tl2);
    \coordinate (dummycoord5) at ($(finaltimestep.east)$) edge[mid] (finaltimestep);
    \draw [mid,draw] (finaltimestep) -| (dummycoord5) node[near end, anchor= south west] {no};
    \node (dummynode3) [below = 1cm of finaltimestep] {} edge[pre] (finaltimestep);
    \coordinate (dummycoord6) at ($(finaltimestep.south)$) edge[mid] (finaltimestep);
    \draw [mid,draw] (dummycoord6) -| (dummynode3) node[near end, anchor= south west] {yes};

\node (exit) [below = 1cm of finaltimestep] {Stop the simulation and exit};

\begin{scope}[on background layer]

\coordinate (southendtimeloopbox) at ($(finaltimestep.south)-(0,10pt)$) (finaltimestep);
\coordinate (northendtimeloopbox) at ($(tl2.north)+(0,2.5pt)$) (tl2);
\coordinate (westendtimeloopbox) at ($(momentum.west)-(2.25cm,0)$) (momentum);
\node [draw=black,rectangle,fill=Azure1,fit= (westendtimeloopbox) (northendtimeloopbox) (tupdate) (southendtimeloopbox), inner sep = 5pt] (timeloop) {};
\node [title, anchor=north west] at ($(timeloop.north west)+(5pt,-5pt)$) {Time loop};

\coordinate (southendinitloopbox) at ($(finalinnerloop.south)-(0,10pt)$) (finalinnerloop);
\coordinate (northendinitloopbox) at ($(innerit2aoihf.north)+(0,2.5pt)$) (innerit2aoihf);
\coordinate (westendinitloopbox) at ($(momentum.west)-(2.0cm,0)$) (momentum);
\node [draw=black,rectangle,fill=Azure2,fit= (westendinitloopbox) (northendinitloopbox) (iupdate) (southendinitloopbox), inner sep = 5pt] (initer) {};
\node [title, anchor=north west] at ($(initer.north west)+(5pt,-5pt)$) {Inner iterations loop};

\coordinate (southendSIMPLECloopbox) at ($(ucorrection.south)-(0,1.5pt)$) (ucorrection);
\coordinate (northendSIMPLECloopbox) at ($(innerit3.north)+(0,2.5pt)$) (innerit3);
\coordinate (westendnSIMPLECoopbox) at ($(ucorrection.west)-(1.75cm,0)$) (ucorrection);
\coordinate (eastendnSIMPLECoopbox) at ($(jupdate.east)+(0.25cm,0)$) (jupdate);
\node [draw=black,rectangle,fill=LightSteelBlue2,fit= (westendnSIMPLECoopbox) (northendSIMPLECloopbox) (eastendnSIMPLECoopbox) (southendSIMPLECloopbox), inner sep = 5pt] (SIMPLEC) {};
\node [title, anchor=north west] at ($(SIMPLEC.north west)+(5pt,-5pt)$) {SIMPLEC};

\coordinate (southendnonorthloopbox) at ($(finalNOLoop.south)-(0,10pt)$) (finalNOLoop);
\coordinate (northendnonorthloopbox) at ($(innerit4.north)+(0,2.5pt)$) (innerit4);
\coordinate (westendnonorthloopbox) at ($(continuity.west)-(1.5cm,0)$) (continuity);
\node [draw=black,rectangle,fill=LightSteelBlue3,fit= (westendnonorthloopbox) (northendnonorthloopbox) (jupdate) (southendnonorthloopbox), inner sep = 5pt] (nonorth) {};
\node [title, anchor=north west] at ($(nonorth.north west)+(5pt,-5pt)$) {Non-orthogonal correction loop};

\end{scope}

\end{tikzpicture}
}
\selectcolormodel{rgb}
\caption{Solver flowchart.}
\label{fig:solver_flowchart}
\end{figure*}
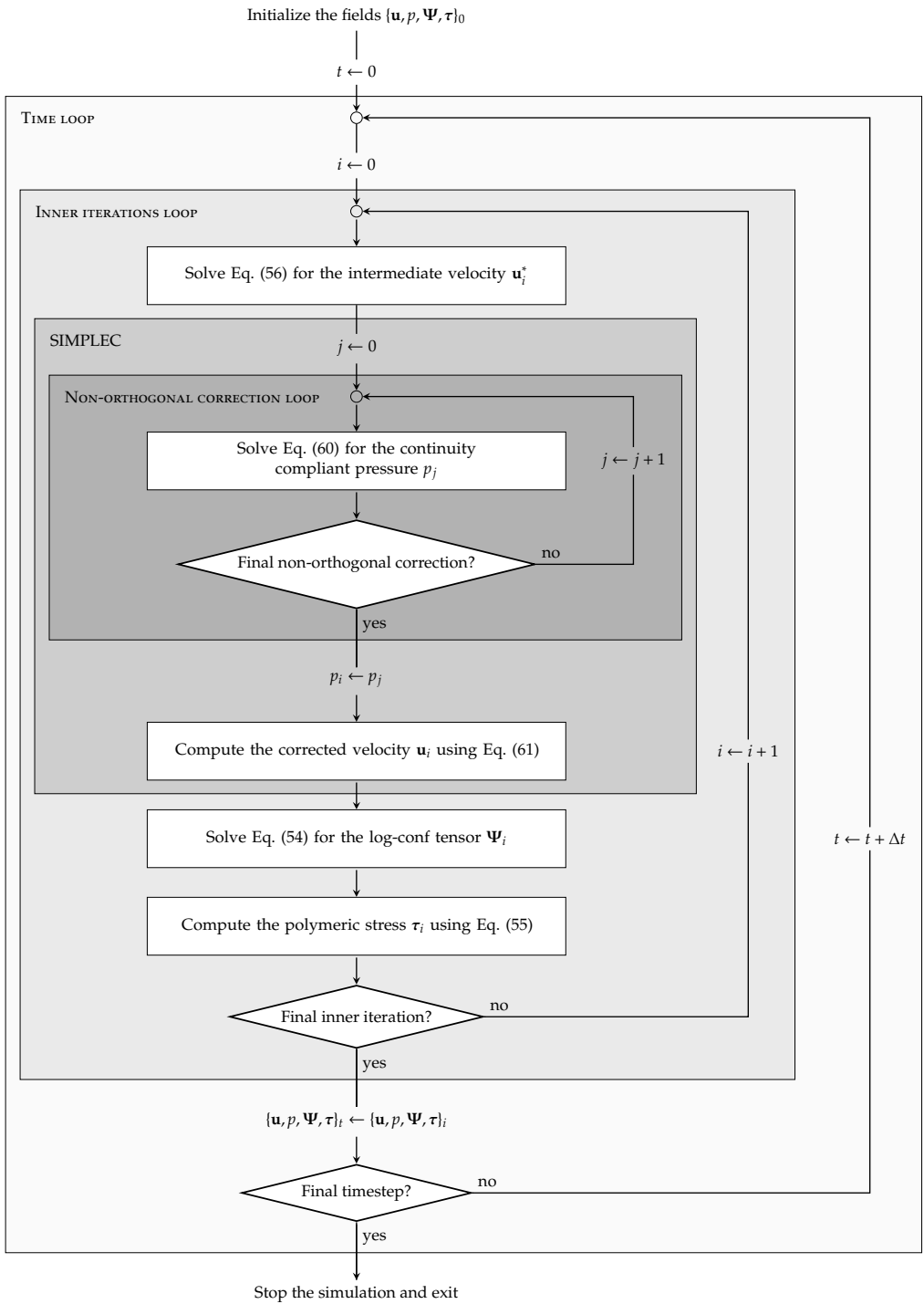

In principle, we now have arrived at a set of decoupled partial differential equations~\eqref{eqn:momentum_ah},\eqref{eqn:pressure_correction} and \eqref{eqn:log-conf_i} and two algebraic evaluations~\eqref{eqn:psi-to-tau_i} and \eqref{eqn:velocity_correction} that can be composed into an algorithm as illustrated in Fig.~\ref{fig:solver_flowchart}.
However, Fig.~\ref{fig:solver_flowchart} contains another interior fixed-point loop around the pressure correction equation~\eqref{eqn:pressure_correction}. The rationale here is that the spatial discretization of the surface gradient $\nabla p_i$, which will be described in the following section, is defective for non-orthogonal meshes. To correct for this, some computations in the scheme are deferred in a non-linear fashion, which then necessitate another fixed-point loop around the discretized version of Eq.~\eqref{eqn:pressure_correction}. The latter happens even though Eq.~\eqref{eqn:pressure_correction} looks linear on the current level of abstraction. For the details, we refer the reader to~\cite[Sec.~9.8]{ferziger2019computational}.

In all simulations that are presented in Section~\ref{sec:benchmarks}, a total of two inner iteration loops and two non-orthogonal correction steps per time-step are used.

\subsection{Spatial discretization}\label{sec:FVMdiscretization}
After temporal discretization, linearization and decoupling of velocity and pressure with the SIMPLEC method, we arrive at three decoupled, linear partial differential equations~\eqref{eqn:momentum_ah},\eqref{eqn:pressure_correction} and \eqref{eqn:log-conf_i}. In order to solve those, we need to choose a method for spatial discretization. As noted earlier, we have chosen the Finite Volume Method (FVM), and in particular base our implementation on RheoTool~\cite{rheoTool} and OpenFOAM\textsuperscript{\textregistered}~\cite{Weller1998}.

In the FVM, the computational domain is subdivided into a set of appropriate interconnected control volumes (the mesh) and the integral form of these PDEs is then evaluated on every single control volume~\cite{ferziger2019computational}. The variables of interest ($\mathbf{u}^*$, $p_i$ and $\mathbf{\Psi}_i$) are, in our choice of a cell-centered FVM, considered as discrete fields (vector-, scalar- and tensorfields, respectively) which attain their respective value at the cell center. The appearing spatial differential operators are then approximated using different schemes that solely depend on those cell-centered quantities. With the initial PDEs being linear, this approach results in sparse linear equation systems.

Next, we list the configuration of the spatial discretization schemes, which will be used throughout all simulations that follow in Section~\ref{sec:benchmarks}.
\begin{itemize}
    \item The divergence terms are discretized according to the divergence theorem via the Gauss scheme. For that, the argument of the divergence operator needs to be evaluated on the faces of the cell. For $\nabla \cdot \bm{\tau}_{i-1}$ or $\nabla \cdot \left(A^{-1} (H\mathbf{u}^* + \mathbf{b})\right)$ this means that the cell-centered value is interpolated linearly from cell to face. In Eq.~\eqref{eqn:pressure_correction} the term $(A-H_1)^{-1} - A^{-1}$ is also linearly interpolated from cell to face.
    \item The Laplacian terms, such as $\nabla \cdot (\eta_s \nabla \mathbf{u}_i)$ and $\nabla \cdot \left(\left(A-H_1\right)^{-1} \nabla p_i\right)$, are also discretized using Gaussian integration, with the difference that only the inner factors are linearly interpolated. The gradients $\nabla \mathbf{u}_i$ and $\nabla p_i$, but also $\nabla p^*$ in Eq.~\eqref{eqn:pressure_correction}, are directly evaluated on the face using a surface normal scheme. In all our computations we have employed a surface normal gradient scheme with an explicit deferred non-orthogonal correction.
    \item Cell-centered gradients, as $\nabla p^*$ and $\nabla \mathbf{u}_i$ in Eqs.~\eqref{eqn:log-conf_i},\eqref{eqn:momentum_ah},\eqref{eqn:velocity_correction}, are computed using the Gauss scheme with linear interpolation. Interpolation in general is linear per default, whenever needed.
    \item For the convective term in the constitutive equation, $(\mathbf{u}_i \cdot \nabla)\mathbf{\Psi}_i$, the corrected, component-wise CUBISTA scheme is used, which is described in~\cite{pimenta2017stabilization}. The convective term $(\mathbf{u}_{i-1} \cdot \nabla) \mathbf{u}_i$ in the momentum balance is removed from Eq.~\eqref{eqn:momentum_ah} in the later benchmarks (to enforce $\text{Re} = 0$) and, therefore, no discretization scheme is needed.
\end{itemize}
Overall, all used spatial discretization schemes are under ideal conditions, i.e., on orthogonal meshes, second order accurate. However, as for example shown in~\cite{Syrakos_2017}, the gradient computation may lose its second order accuracy on meshes of poor quality, e.g., high non-orthogonalities or skewnesses. As a consequence, particular attention was paid to the selection and design of the hexahedral meshes in Section~\ref{sec:benchmarks}.

A crucial aspect when simulating incompressible Navier--Stokes equations, regardless of the actually employed spatial discretization scheme, are the issues with checkerboard patterns and in general the saddle-point structure of the linearized problem. Here, this issue has been approached with the Rhie--Chow method~\cite{rhie1983numerical}, where $\nabla p_i, \nabla p^*$ are differently discretized in Eq.~\eqref{eqn:pressure_correction} than they are in Eqs.~\eqref{eqn:momentum_ah} and~\eqref{eqn:velocity_correction}. We do not want to go into the details here, since they have already been laid out in~\cite{pimenta2017stabilization}, but solely mention two points: Firstly, there is a connection to the---in the finite element world important---inf-sup condition, and we refer the interested reader to~\cite{negrini2023rhie} for a recent account into that direction. Secondly, on top of what has just been described, OpenFOAM\textsuperscript{\textregistered} employs a correction of the flux in Eq.~\eqref{eqn:pressure_correction} that shall remedy unphysical dependencies of steady-state solutions on the actually chosen time-step size. The reader is once again referred to~\cite{pimenta2017stabilization} for the details.

Of course, boundary conditions do also constitute an important aspect of numerical methods for partial differential equations. The specific choice of boundary conditions for the later benchmarks will follow in the corresponding sections~\ref{sec:confinedCylinder} and~\ref{sec:sedimentingSphere}. Nonetheless, it should be noted that boundary conditions are handled according to the technique that is implemented in OpenFOAM\textsuperscript{\textregistered}, where specific boundary structures, called patches, are used to store boundary information. Hence, whenever needed by a certain discretization scheme for elements at the edge of the computational domain, the required values, that cannot be provided by interior neighbors, are fetched form these boundary patches.

Finally, we will mention that the choice of the viscoelastic model (e.g., Oldroyd-B or Giesekus) and in particular the implementation of the eigenvalue-free $f(\opad\mathbf{\Psi})$ term does not affect the overall procedure depicted in Fig.~\ref{fig:solver_flowchart}, but rather the assembling of the right-hand side of Eq.~\eqref{eqn:log-conf_i}. It can therefore be implemented quite straightforwardly as described in Section~\ref{sec:eigVal-Free_Algo} by computing the $f(\opad\mathbf{\Psi})$ term according to Algorithm~\ref{alg:fopad}.

\subsection{Choice of linear solvers} \label{sec:linear_solvers}
Through the spatial discretization in the last section, we now have effectively derived three systems of sparse linear equation system that correspond to Eqs.~\eqref{eqn:momentum_ah},\eqref{eqn:pressure_correction},\eqref{eqn:log-conf_i} and which are solved for the cell-centered values of $\mathbf{u}^*, p_i$ and $\mathbf{\Psi}_i$. For the rest of this section, we will refer to these systems as the $\mathbf{u}^*, p_i$ and $\mathbf{\Psi}_i$ equation respectively. One immediate computational optimization, which is employed in OpenFOAM\textsuperscript{\textregistered}, is that the left-hand sides of Eqs.~\eqref{eqn:pressure_correction} and~\eqref{eqn:log-conf_i} can be decoupled and solved individually for the components of $\mathbf{u}^*,\mathbf{\Psi}_i$.

After this optimization, the individual linear systems are solved using the following solvers:
For the $\mathbf{u}^*$ and $p_i$ equations, the Preconditioned Conjugate Gradient Method (PCG) is applied with an Diagonal-Based Incomplete Cholesky preconditioner (DIC). An absolute tolerance of $10^{-10}$, relative tolerance of $10^{-4}$ and a maximum number of 1000 iterations are chosen as the possible termination criteria for these solvers. The $\mathbf{\Psi}_i$ equation uses a Preconditioned Bi-Conjugate Gradient method (PBiCG) with an Diagonal-Based Incomplete LU preconditioner (DILU). The same termination configuration is chosen as for the $\mathbf{u}^*$ and $p_i$ equations.

In our numerical algorithm, the currently available field data is used as the initial guess for the corresponding iterative solver. In our benchmarks, a dimensionless timescale $T = t/\lambda$ is used and each simulation is run until $T = 30$ with a Courant number of 0.5. We, therefore, ensure that the viscoelastic stresses in the fluid have converged at the end of a simulation, i.e., that the fluid has reached a steady-state. Within this steady-state, the initial guesses for the iterative solvers will already be close to the actual solutions, such that the number of iterations is expected to decrease as the simulation progresses in time. However, in a non-steady-state, i.e., at the beginning of a simulation, the initial guesses may be quite far from the actual solution of the system, such that more iterations are needed in general.

Typically, the $p_i$ equation is the most expensive to solve. At the beginning of a simulation, the $p_i$ equation requires several hundred iterations for convergence or even reaches the maximum number of iterations on our finest meshes. Overall the number of iterations needed for convergence decreases as the fluid approaches a steady-state. In a steady-state, there is often no need for a single iteration of the $\mathbf{u}^*$ and $\mathbf{\Psi}_i$ equations, since the initial guess already solves the system well enough.

\section{Benchmarks}
\label{sec:benchmarks}
In this section, our implementation of the newly derived eigenvalue-free constitutive formulation is applied to a study of two benchmarks: the confined cylinder and the sedimenting sphere.
These benchmarks represent similar flow problems, i.e., flow around an obstacle, in a two-dimensional and a three-dimensional case, respectively. 
Both benchmarks have been examined in the literature before, in order to validate new numerical schemes or models, see for example~\cite{Knechtges2014, Hulsen2005, Claus2013, Fan1999, Liu1998, Sun1999, Afonso2009} for the confined cylinder and~\cite{Knechtges2015, Lunsmann1993, Owens1996, Chauviere2000, Fan2003} for the sedimenting sphere.
For comparability, we specifically follow the setups, i.e., the geometries and fluid parameters, that were used in~\cite{Knechtges2014} for the confined cylinder and~\cite{Knechtges2015} for the sedimenting sphere. A detailed description will follow in the corresponding sections~\ref{sec:confinedCylinder} and \ref{sec:sedimentingSphere}, where results for the eigenvalue-free logarithmic Oldroyd-B and Giesekus models are shown and discussed.

The main quantity of interest in both benchmarks is the drag coefficient $C_d$, which describes the non-dimensionalized force the fluid exerts on the obstacle in $x$-direction. $C_d$ is given by
\begin{equation}
    C_d = \frac{1}{(\eta_s+\eta_p)\bar{u}} \int_{\Gamma} \mathbf{e}_x \cdot (\bm{\sigma}\mathbf{n}) \, , 
    \label{eqn:cd}
\end{equation}
where $\Gamma$ is the surface of the obstacle, $\mathbf{n}$ the corresponding unit normal, $\mathbf{e}_x$ the unit vector in $x$-direction and $\bm{\sigma}$ the Cauchy stress tensor
\begin{equation}
    \bm{\sigma} = -p\mathbf{1} + \eta_s(\nabla\mathbf{u}+\nabla\mathbf{u}^T) + \bm{\tau} \, .
    \label{eqn:sigma}
\end{equation}
It is known that the drag coefficient varies with the Reynolds number Re of the simulation. This has for example been investigated by~\cite{Claus2013}. However, for comparability, we follow the literature and consider creeping flow conditions ($\text{Re}=0$) in both benchmarks by removing the convective term from the momentum equation~\eqref{eqn:momentum}.

Overall, a variety of flow simulations for different Weissenberg numbers will be presented and the corresponding drag coefficient values will be compared to the literature. The dimensionless Weissenberg number is given by
\begin{equation}
    \text{Wi} = \frac{\lambda \bar{u}}{R} \, ,
    \label{eqn:Wi}
\end{equation}
where $\lambda$ is the relaxation time of the fluid, $R$ is the radius of the cylinder or the sphere and $\bar{u}$ the mean inflow velocity.

\subsection{Confined cylinder}
\label{sec:confinedCylinder}
\begin{figure}[t!]
    \centering
    \includegraphics[width=\linewidth]{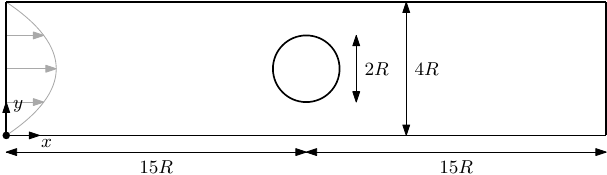}
    \caption{Illustration (not to scale) of the confined cylinder. Fluid flows from the inlet at the left side to the outlet at the right side. The upper and lower boundaries of the channel and the cylinder surface are considered as solid walls.}
    \label{fig:confinedCylinder}
\end{figure}

In the confined cylinder case, a two-dimensional channel with a cylindrical obstacle of radius $R$ in its center is considered as the computational domain. The channel has a height of $4R$, such that the ratio of the channel height to the cylinder diameter is 2. Our setup mimics the setup of Knechtges~et~al.~\cite{Knechtges2014} and Hulsen~et~al.~\cite{Hulsen2005}, where the channel has a total length of $30R$ in order to reduce effects of the inflow and outflow and where the cylinder center is at $(15R,2R)$.
An illustration of the geometry can be seen in Fig.~\ref{fig:confinedCylinder}.

\begin{figure}[h]
    \centering
    \import{}{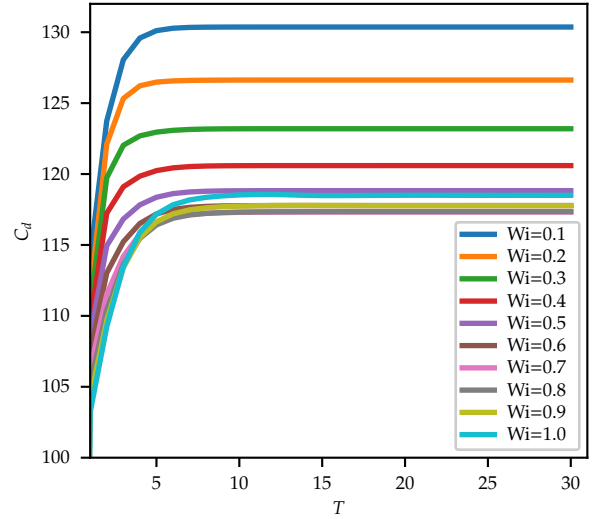}
    \caption{Convergence of the $C_d$ values for the confined cylinder case on mesh M3 at different Weissenberg numbers over time from $T=1$ to $T=30$ using the eigenvalue-free logarithmic Oldroyd-B formulation.}
    \label{fig:Cd_OBLog_confCyl}
\end{figure}

\begin{figure*}[t!]
    \centering
    \includegraphics[width=\textwidth]{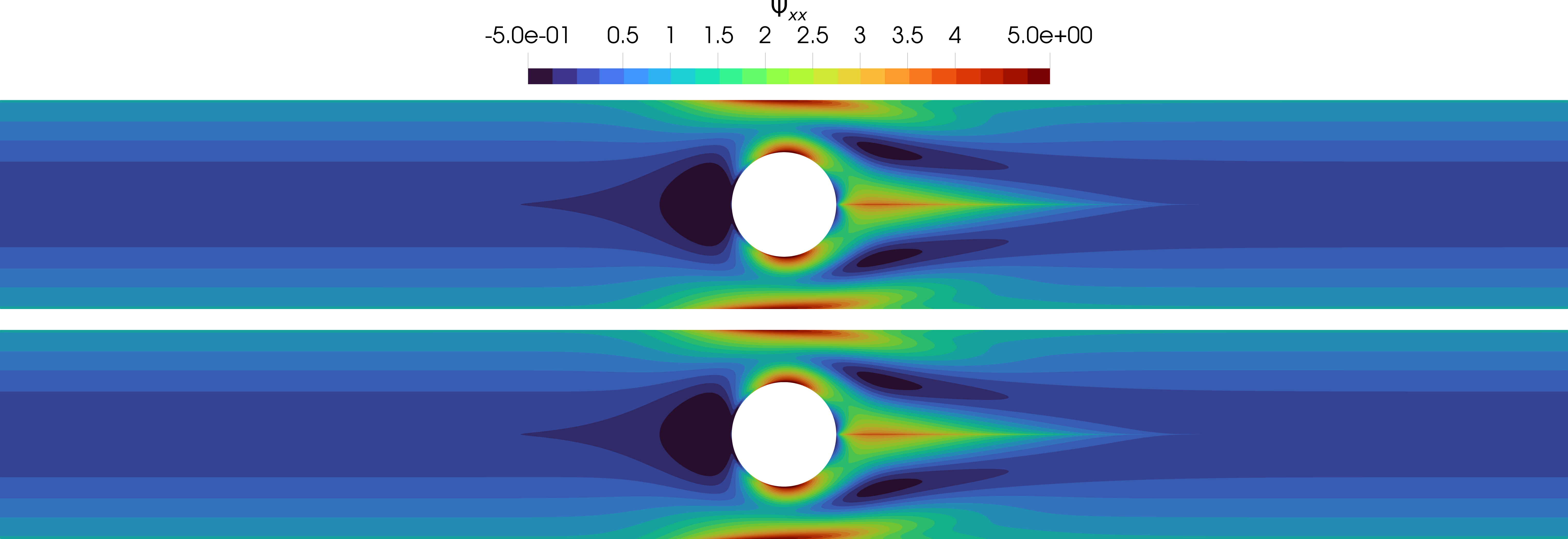}
    \caption{Comparison of $\Psi_{xx}$ at the final time-step $T=30$. Top: computed with the eigenvalue-free logarithmic Oldroyd-B formulation; bottom: computed with the standard logarithmic Oldroyd-B formulation that relies on an eigenvalue decomposition. Looking at the entrance of both simulations, it can additionally be seen that developed Poiseuille inflow conditions have been used, since the $\Psi_{xx}$ components are already developed at the inlet.}
    \label{fig:eigValFree_tauxx_comparison_Wi07}
\end{figure*}

\begin{table*}[t!]
    \centering
    \setlength\extrarowheight{2pt}
    \begin{tabular*}{\textwidth}{@{\extracolsep{\fill}}*{1}{l}@{\extracolsep{\fill}}*{3}{c}}
    \hline
     & M1 & M2 & M3 \\
    \hline
    Number of elements in the mesh & 99576 & 398304 & 1593216\\ 
    Number of elements on the cylinder surface & 756 & 1512 & 3024\\
    Average element non-orthogonality & 12.6 & 12.6 & 12.7\\
    Maximum element non-orthogonality & 44.7 & 44.9 & 45.0\\
    Maximum element skewness & 1.5 & 1.5 & 1.5 \\
    \hline
    \end{tabular*}
    \caption{Mesh statistics for the confined cylinder geometry.}
    \label{tab:meshes_confinedCylinder}
\end{table*}

\subsubsection{Setup}
Boundary and initial conditions are chosen according to literature. At the inlet, a fully developed Poiseuille solution for an Oldroyd-B fluid is imposed for the velocity $\mathbf{u}$ (with mean inflow $\bar{u}$) and the polymeric extra stresses $\bm{\tau}$ and $\mathbf{\Psi}$, similar to~\cite{Knechtges2014}. A zero-gradient condition is considered for the pressure $p$. The exact values for the Poiseuille flow are given in \ref{sec:PoiseuilleInflowConditions}. Furthermore, at the channel and cylinder walls, zero-gradient conditions are considered for the pressure and zero velocities ($\mathbf{u} = \mathbf{0}$). The polymeric extra stress components are linearly extrapolated. At the outlet, zero-gradient conditions are imposed for all variables, except for the pressure, which is set to zero. Initially ($t=0$) the fluid is at rest ($\mathbf{u} = \mathbf{0}$) and the extra-stresses are null ($\bm{\tau} = \mathbf{\Psi} = \mathbf{0}$). The pressure is set to zero as well.

In all of the following tests, $R=\SI{1}{m}$ and $\bar{u}=\SI{1}{m/s}$ were fixed, such that the Weissenberg number equals the numerical value of the relaxation time in seconds and could therefore easily be controlled by a change of $\lambda$. Finally, as in the corresponding literature, a viscosity ratio of $\beta=\eta_s/(\eta_s+\eta_p)=0.59$ and a density of $\rho=\SI{1}{kg/m^3}$ have been used.\footnote{The parameters for our tests are chosen according to the literature for comparability and do not represent real fluids.}

Three quadrilateral meshes M1, M2 and M3 of different refinement levels have been considered. Their main properties are shown in Tab.~\ref{tab:meshes_confinedCylinder}. In each refinement step the total number of elements is quadrupled from mesh to mesh and the number of elements at the cylinder surface is doubled. An important property of these meshes and their refinement is that characteristics, such as the element non-orthogonality and skewness, are sufficiently small. Element non-orthogonality refers to the angle between the vector of two neighbored cell centers and their corresponding face normal. Element skewness refers to the deviation of the intersection point of this cell-center-connecting vector from the actual face center. For example, in a pure square mesh, element non-orthogonality and skewness would both be zero. In the FVM, the gradient computation can be negatively affected by such mesh irregularities, as is described and investigated by Syrakos~et~al.~\cite{Syrakos_2017}. Furthermore, the importance of good quality meshes and strategic mesh refinement is particularly emphasized in~\cite{ferziger2019computational}. Therefore, only mesh configurations were considered where these characteristic values were sufficiently small on all refinement levels. RheoTool does already provide a confined cylinder case with an appropriate mesh~\cite{rheoTool}. The latter has been used as the basis for our benchmarks and adjusted, e.g., by adding several different refinement levels for the mesh. It should also be mentioned, that in order to achieve reasonable $C_d$ values, boundary layers around the obstacle surface were used. This use of thin boundary layers has increased the resolution of the solution close to the obstacle surface and did also reduce the extrapolation error, resulting in $C_d$ values that are in good agreement with the literature.

As already mentioned in Section~\ref{sec:linear_solvers}, adaptive time-stepping kept a Courant number of 0.5 in all simulations. Typical time-step sizes were then ranging from $\SI{1.8e-3}{s}$ on M1, to $\SI{9.0e-4}{s}$ on M2, and $\SI{4.5e-4}{s}$ on M3. For all simulations, a dimensionless timescale $T = t/\lambda$ was used with end time $T=30$ in order to ensure convergence of the fluid to a steady-state. Therefore, the $C_d$ values also converge eventually, as can be seen in Fig.~\ref{fig:Cd_OBLog_confCyl}.

\subsubsection{Results}
Tab.~\ref{tab:Cd_values_confinedCylinder_OBLog} shows the final $C_d$ values for the eigenvalue-free logarithmic Oldroyd-B formulation.
\begin{table*}[h!]
    \centering
    \setlength\extrarowheight{2pt}
    \begin{tabular*}{\textwidth}{@{\extracolsep{\fill}}*{1}{l}@{\extracolsep{\fill}}*{7}{c}}
    \hline
    \multirow{2}{*}{Wi} & \multicolumn{7}{c}{$C_d$}\\
    \cline{2-8}
     & M1 & M2 & M3 & \cite{Knechtges2014} & \cite{Hulsen2005} & \cite{Claus2013} & \cite{Fan1999}\\
    \hline
    0.1 & 130.31898 & 130.36049 & 130.36653 & 130.3626 & 130.363 & 130.364 & 130.36\\
    0.2 & 126.58894 & 126.62264 & 126.62875 & 126.6252 & 126.626 & 126.626 & 126.62\\
    0.3 & 123.16959 & 123.18940 & 123.19475 & 123.1912 & 123.193 & 123.192 & 123.19\\
    0.4 & 120.59084 & 120.59124 & 120.59500 & 120.5912 & 120.596 & 120.593 & 120.59\\
    0.5 & 118.85227 & 118.82872 & 118.83021 & 118.8260 & 118.836 & 118.826 & 118.83\\
    0.6 & 117.83174 & 117.78125 & 117.77988 & 117.7752 & 117.775 & 117.776 & 117.78\\
    0.7 & 117.40242 & 117.32483 & 117.32079 & 117.3157 & 117.315 & 117.316 & 117.32\\
    0.8 & 117.45188 & 117.35293 & 117.35114 & 117.3454 & 117.373 & 117.368 & 117.36\\
    0.9 & 117.87883 & 117.76574 & 117.77477 & 117.7678 & 117.787 & 117.812 & 117.80\\
    1.0 & 118.60224 & 118.47727 & 118.49927 &          & 118.471 &         & 118.49\\
    \hline
    \end{tabular*}
    \caption{Final values for the drag coefficient $C_d$ at $T=30$ for the confined cylinder case, using the eigenvalue-free Oldroyd-B formulation at different Weissenberg numbers.}
    \label{tab:Cd_values_confinedCylinder_OBLog}
\end{table*}
Overall, the results on the finest mesh M3 show good agreement with the literature at all considered Weissenberg numbers. At smaller Weissenberg numbers ($\text{Wi}\leq0.7$) the values in the compared publications~\cite{Knechtges2014, Hulsen2005, Claus2013, Fan1999} deviate at a magnitude of $10^{-3}$ and our results on M3 (which we consider as our most accurate ones) do also fit into this range. At higher Weissenberg numbers the values tend to deviate more from each other among all publications, roughly at a magnitude of $10^{-2}$; a property that has already been observed and described for example in~\cite{Knechtges2014}. Furthermore, all publications agree that the minimum drag coefficient is obtained at $\text{Wi} = 0.7$. The highest $C_d$ values of around 130.36 are reached at the lowest Weissenberg number of 0.1.

Fig.~\ref{fig:eigValFree_tauxx_comparison_Wi07} shows solutions of the confined cylinder case  at $T=30$ and a Weissenberg number $\text{Wi}=0.7$. Presented are the $\Psi_{xx}$ components for the eigenvalue-free logarithmic Oldroyd-B formulation in comparison with a eigenvalue-based formulation, that is described by Pimenta~\cite[Eq. (7)]{pimenta2017stabilization} and previously implemented in RheoTool~\cite{rheoTool}. The contours of the tensor components, and in particular those close to the cylinder, look almost identical. To emphasize and quantify the similarity of these solutions, it can additionally be stated that their final $C_d$ difference is only of magnitude $10^{-6}$.

Tab.~\ref{tab:Cd_values_confinedCylinder_GiesekusLog} shows $C_d$ results for computations with the eigenvalue-free logarithmic Giesekus model.
\begin{sidewaystable*}
    \centering
    \setlength\extrarowheight{2pt}
    \begin{tabular*}{\textwidth}{@{\extracolsep{\fill}}*{1}{l}@{\extracolsep{\fill}}*{12}{c}}
    \hline
    \multirow{3}{*}{Wi} & \multicolumn{12}{c}{$C_d$} \\
    \cline{2-13}
    & \multicolumn{4}{c}{$\alpha = 0.1$} & \multicolumn{4}{c}{$\alpha = 0.01$} & \multicolumn{4}{c}{$\alpha = 0.001$}\\
    \cline{2-5} \cline{6-9} \cline{10-13}
    & M1 & M2 & M3 & \cite{Claus2013} & M1 & M2 & M3 & \cite{Claus2013} & M1 & M2 & M3 & \cite{Claus2013}\\
    \hline
    0.1 & 125.542 & 125.585 & 125.591 & 125.587 & 129.626 & 129.667 & 129.674 & 129.671 & 130.246 & 130.287 & 130.293 & 130.291\\
    0.2 & 117.068 & 117.109 & 117.116 & 117.113 & 124.629 & 124.666 & 124.672 & 124.670 & 126.358 & 126.392 & 126.398 & 126.396\\
    0.3 & 111.055 & 111.095 & 111.102 & 111.098 & 120.050 & 120.081 & 120.087 & 120.085 & 122.753 & 122.775 & 122.780 & 122.778\\
    0.4 & 106.814 & 106.852 & 106.859 & 106.855 & 116.487 & 116.513 & 116.519 & 116.517 & 119.974 & 119.979 & 119.984 & 119.981\\
    0.5 & 103.694 & 103.731 & 103.737 & 103.733 & 113.842 & 113.863 & 113.869 & 113.867 & 118.020 & 118.005 & 118.008 & 118.005\\
    0.6 & 101.304 & 101.340 & 101.345 & 101.341 & 111.884 & 111.900 & 111.906 & 111.906 & 116.756 & 116.721 & 116.722 & 116.719\\
    0.7 &  99.413 &  99.447 &  99.452 &  99.448 & 110.401 & 110.415 & 110.421 & 110.422 & 116.040 & 115.986 & 115.985 & 115.982\\
    0.8 &  97.875 &  97.908 &  97.913 &  97.909 & 109.238 & 109.249 & 109.255 & 109.258 & 115.736 & 115.666 & 115.665 & 115.679\\
    0.9 &  96.599 &  96.631 &  96.636 &  96.631 & 108.287 & 108.297 & 108.302 & 108.307 & 115.724 & 115.641 & 115.642 & 115.664\\
    1.0 &  95.520 &  95.552 &  95.556 &  95.552 & 107.483 & 107.491 & 107.496 & 107.505 & 115.907 & 115.811 & 115.813 & 115.868\\
    \hline
    \end{tabular*}
    \caption{Final values for the drag coefficient $C_d$ at $T=30$ for the confined cylinder case, using the eigenvalue-free Giesekus formulation at different Weissenberg numbers. Three different mobility factors $\alpha \in \{0.1, 0.01, 0.001\}$ were considered.}
    \label{tab:Cd_values_confinedCylinder_GiesekusLog}
\end{sidewaystable*}
The Giesekus model has an additional parameter, the mobility factor $\alpha \in [0, 1]$. Again, good agreement with the literature can be observed. Additionally, our results show the significant influence of $\alpha$ on the drag coefficient. We do not want to go into detail here, as the effect of $\alpha$ on $C_d$ has already been investigated by others, see for example~\cite{Claus2013}. As $\alpha$ increases (for fixed Wi), the drag decreases, which is explained by the shear-thinning property of the Giesekus model. When $\alpha$ tends to zero, the Giesekus model transitions to the Oldroyd-B model and thus, the $C_d$ values converge to the corresponding values in Tab.~\ref{tab:Cd_values_confinedCylinder_OBLog}.

\begin{figure}[h!]
    \centering
    \includegraphics[width=\linewidth]{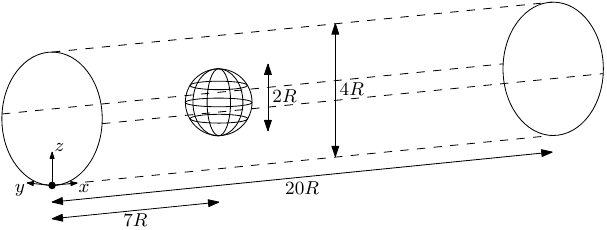}
    \caption{Illustration (not to scale) of the sedimenting sphere. Fluid flows from the inlet at the left side to the outlet at the right side. A fixed non-zero velocity is considered at the channel wall. A sphere with solid surface (zero velocity) is placed inside the channel.}
    \label{fig:sedimentingSphere}
\end{figure}

\begin{figure*}[t!]
    \centering
    \includegraphics[width=.7\linewidth]{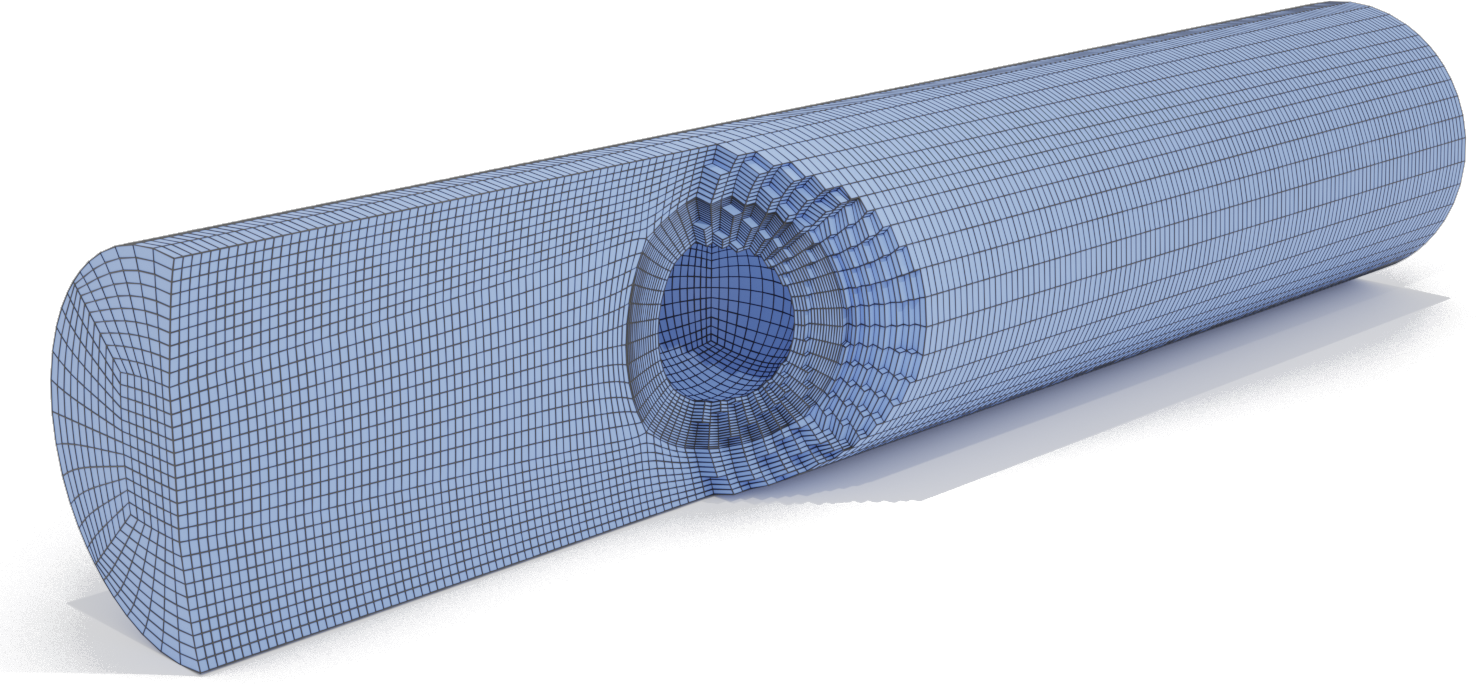}
    \caption{Rendering of the hexahedral mesh M1 for the sedimenting sphere benchmark.}
    \label{fig:sedSphere_M1_variant3p}
\end{figure*}

All computations were run in parallel on the Caro HPC cluster of the German Aerospace Center. M1 simulations were run on 32 cores, M2 simulations on 64 cores and M3 simulations on 128 cores. In its current state, we observe that our implementation of the eigenvalue-free variant is slightly slower than the standard eigenvalue-based implementation. In particular, we measure a runtime increase of around 7\% per time-step in the log-conf equation. However, solving the constitutive equation for a single relaxation mode has only a minor impact on the overall runtime of the algorithm, since the momentum equation and the SIMPLEC algorithm are more computationally heavy. This is corroborated by the comparison of the total runtimes for our test case on M3 using the eigenvalue-free formulation with those of the standard formulation, which differ by less than 1\%. 

Further performance optimizations of the Algorithms~\ref{alg:fopad} and~\ref{alg:hx} are possible, but at the moment not considered in our prototypical implementation. At the moment, e.g., our implementation does not exploit the matrix symmetry of $\mathbf{X}$ in Algorithm~\ref{alg:hx}, which could easily save some floating point operations. Another optimization opportunity that is currently unexploited, and which is for the eigenvalue-based implementations much more difficult to pursue, is to bring the actual computations onto a GPU.

We also applied the eigenvalue-free approach to other simulation cases at higher Weissenberg numbers and did not observe any significant differences regarding its stability compared to the standard approach.

\subsection{Sedimenting sphere}
\label{sec:sedimentingSphere}
To demonstrate the eigenvalue-free approach on a three-dimensional problem, a simulation similar to the confined cylinder, the sedimenting sphere, is considered. In this benchmark, fluid flow around a spherical obstacle inside a three-dimensional channel is considered. The sphere has a radius of $R$ and the channel a height (or diameter) of $4R$. Based on the setup in~\cite{Knechtges2015}, we impose a channel length of $20R$ and keep the sphere centered at $(7R, 0, 2R)$. An excerpt of the computational domain is shown in Fig.~\ref{fig:sedimentingSphere}.

\begin{table*}[t!]
    \centering
    \setlength\extrarowheight{2pt}
    \begin{tabular*}{\textwidth}{@{\extracolsep{\fill}}*{1}{l}@{\extracolsep{\fill}}*{3}{c}}
    \hline
     & M1 & M2 & M3 \\
    \hline
    Number of elements in the mesh & 139392 & 1115136 & 8921088\\ 
    Number of elements on the sphere surface & 1152 & 4608 & 18432\\
    Average element non-orthogonality & 11.1 & 11.7 & 12.0\\
    Maximum element non-orthogonality & 41.1 & 52.4 & 64.4\\
    Maximum element skewness & 1.7 & 1.8 & 1.8\\
    \hline
    \end{tabular*}
    \caption{Mesh statistics for the sedimenting sphere geometry.}
    \label{tab:meshes_sedimentingSphere}
\end{table*}

\subsubsection{Setup}
Boundary and initial conditions are chosen according to the literature in order to increase comparability. A uniform inlet condition is considered, with a fixed non-zero velocity $\bar{u}$ in $x$-direction, zero polymeric extra stress components, and a zero-gradient condition for the pressure. At the channel wall, the boundary conditions are chosen equal to the inlet conditions. Thus, in particular, the velocity is uniformly fixed with non-zero component in $x$-direction as well. At the sphere, a no-slip condition for the velocity is considered ($\mathbf{u} = \mathbf{0}$). The polymeric extra stress components are linearly extrapolated onto the surface and the pressure uses a zero-gradient condition. At the outlet, zero-gradient conditions are imposed for all variables except for the pressure, which uses a fixed value condition $p=0$.

In the following tests, $R=\SI{1}{m}$ and $\bar{u}=\SI{1}{m/s}$ were used, such that the Weissenberg number equals the numerical value of the relaxation time in seconds and can again be controlled by a change of $\lambda$. As in the corresponding literature, a viscosity ratio of $\beta=\eta_s/(\eta_s+\eta_p)=0.5$ and a density of $\rho=\SI{1}{kg/m^3}$ have been used.

Three purely hexahedral meshes M1, M2 and M3 of different refinement levels have been considered. Their main properties are shown in Tab.~\ref{tab:meshes_sedimentingSphere}. During refinement, the total number of elements is multiplied by eight from mesh to mesh, while the number of elements at the sphere surface is quadrupled. It was observed that the $C_d$ computation in this case was very sensitive to the overall mesh quality. Configuring the mesh for the sedimenting sphere simulations, with the goal to minimize non-orthogonalities and skewnesses on all refinement levels, did therefore play an important role during our research. Additionally, boundary layers around the sphere surface were introduced for smaller numerical errors close to the surface and, therefore, a better $C_d$ accuracy. An excerpt of mesh M1 is presented in Fig.~\ref{fig:sedSphere_M1_variant3p}. Again, an end time $T=30$ was used and a Courant number of 0.5 was fixed, leading to typical time-step sizes ranging from $\SI{1.6e-2}{s}$ on M1, to $\SI{8.0e-3}{s}$ on M2 and $\SI{4.0e-3}{s}$ on M3. At this point, it should be mentioned that the quantity of interest in the following tests is the drag correction factor $K$, which is typically used in sedimenting sphere benchmarks~\cite{Knechtges2015, Lunsmann1993, Owens1996, Chauviere2000, Fan2003}. $K$ is given by
\begin{equation}
    K = \frac{C_d}{6\pi} \, ,
\end{equation}
where $C_d$ is the drag coefficient value from Eq.~\eqref{eqn:cd} with $\Gamma$ being the sphere surface. The definition of $K$ is motivated by Stokes' law~\cite{stokes1901effect, pau1990application}.

\subsubsection{Results}
Tab.~\ref{tab:Cd_values_sedimentingSphere_OBLog} shows $K$ values for the eigenvalue-free logarithmic Oldroyd-B formulation and varying Weissenberg numbers between 0.1 and 1.5. 
\begin{table*}[h!]
    \centering
    \setlength\extrarowheight{2pt}
    \begin{tabular*}{\textwidth}{@{\extracolsep{\fill}}*{1}{l}@{\extracolsep{\fill}}*{9}{c}}
    \hline
    \multirow{2}{*}{Wi} & \multicolumn{9}{c}{$K$}\\
    \cline{2-10}
     & M1 & M2 & M3 & RE & \cite{Knechtges2015} & \cite{Lunsmann1993} & \cite{Owens1996} & \cite{Chauviere2000} & \cite{Fan2003}\\
    \hline
    0.1 & 5.73784 & 5.82977 & 5.86723 & 5.90469 & 5.90576 &         &        &        & \\
    0.2 & 5.64635 & 5.73532 & 5.77120 & 5.80708 & 5.80763 &         &        &        & \\
    0.3 & 5.53994 & 5.62529 & 5.65930 & 5.69331 & 5.69356 & 5.69368 & 5.6963 &        & \\
    0.4 & 5.43888 & 5.52076 & 5.55300 & 5.58524 & 5.58527 &         &        &        & \\
    0.5 & 5.35026 & 5.42977 & 5.46043 & 5.49109 & 5.49093 &         &        & 5.4852 & \\
    0.6 & 5.27577 & 5.35396 & 5.38330 & 5.41264 & 5.41227 & 5.41225 & 5.4117 & 5.4009 & \\
    0.7 & 5.21468 & 5.29244 & 5.32071 & 5.34898 & 5.34838 &         &        & 5.3411 & \\
    0.8 & 5.16544 & 5.24335 & 5.27092 & 5.29849 & 5.29747 &         &        & 5.2945 & \\
    0.9 & 5.12649 & 5.20481 & 5.23202 & 5.25923 & 5.25761 & 5.25717 &        & 5.2518 & \\
    1.0 & 5.09638 & 5.17511 & 5.20219 & 5.22927 & 5.22700 &         &        & 5.2240 & \\
    1.1 & 5.07430 & 5.15274 & 5.17989 & 5.20704 & 5.20402 &         &        & 5.2029 & \\
    1.2 & 5.05872 & 5.13653 & 5.16379 & 5.19105 & 5.18733 & 5.18648 &        & 5.1842 & 5.1877\\
    1.3 & 5.04914 & 5.12552 & 5.15281 & 5.18010 & 5.17581 &         &        &        & 5.1763\\
    1.4 & 5.04439 & 5.11890 & 5.14608 & 5.17326 & 5.16851 &         &        &        & \\
    1.5 & 5.04361 & 5.11609 & 5.14291 & 5.16973 &         & 5.15293 &        &        & \\
    \hline
    \end{tabular*}
    \caption{Final values for the drag coefficient $C_d$ at $T=30$ for the sedimenting sphere case, using the eigenvalue-free Oldroyd-B formulation at different Weissenberg numbers. RE corresponds to the Richardson extrapolation value.}
    \label{tab:Cd_values_sedimentingSphere_OBLog}
\end{table*}
In most cases, the compared publications show similar values up to a magnitude of $10^{-3}$. In comparison, our results differ slightly more, at a magnitude of $10^{-2}$. However, the mesh convergence of our results suggests that better values could possibly be reached when considering even finer meshes M4, M5 etc. To emphasize this point, we apply a Richardson extrapolation with the discretization length $h$ as a parameter. In our setting, we expect the error in $C_d$ to scale linearly with $h$, since we are using a piecewise linear approximation of the sphere surface when computing the integral in Eq.~\eqref{eqn:cd}. Furthermore, the discretization length $h$ is divided by two in each refinement step. In this case, the Richardson extrapolation value $K_{\text{RE}}$ of the drag correction factor using the obtained values for M2 and M3 yields $K_\text{RE} = 2K_\text{M3} - K_\text{M2}$. The resulting values are shown in the RE column of Tab.~\ref{tab:Cd_values_sedimentingSphere_OBLog} and they show a very good agreement with the compared publications, now deviating at a magnitude of $10^{-3}$ as well. However, as already observed by Knechtges~\cite{Knechtges2015}, the results start to deviate more from each other with increasing Weissenberg numbers, especially for $Wi \ge 1.4$. Finally, it can be noted that all data in Tab.~\ref{tab:Cd_values_sedimentingSphere_OBLog} agrees on the overall trend of decreasing $K$ values for increasing Weissenberg numbers.

Tab.~\ref{tab:Cd_values_sedimentingSphere_GiesekusLog} shows $K$ values for the eigenvalue-free logarithmic Giesekus model, evaluated for the same variety of Weissenberg numbers as before and mobility factors $\alpha \in \{0.1, 0.01, 0.001\}$. 
\begin{sidewaystable*}
    \centering
    \setlength\extrarowheight{2pt}
    \begin{tabular*}{\textwidth}{@{\extracolsep{\fill}}*{1}{c}@{\extracolsep{\fill}}*{12}{c}}
    \hline
    \multirow{3}{*}{Wi} & \multicolumn{12}{c}{$K$} \\
    \cline{2-13}
    & \multicolumn{4}{c}{$\alpha = 0.1$} & \multicolumn{4}{c}{$\alpha = 0.01$} & \multicolumn{4}{c}{$\alpha = 0.001$} \\
    \cline{2-5} \cline{6-9} \cline{10-13}
    & M1 & M2 & M3 & \cite{Knechtges2015} & M1 & M2 & M3 & \cite{Knechtges2015} & M1 & M2 & M3 & \cite{Knechtges2015}\\
    \hline
    0.1 & 5.65893 & 5.74798 & 5.78425 & 5.82166 & 5.72846 & 5.82003 & 5.85734 & 5.89573 & 5.73688 & 5.82878 & 5.86622 & 5.90473\\
    0.2 & 5.42343 & 5.50506 & 5.53791 & 5.57160 & 5.61367 & 5.70137 & 5.73675 & 5.77275 & 5.64289 & 5.73172 & 5.76754 & 5.80393\\
    0.3 & 5.18896 & 5.26348 & 5.29312 & 5.32349 & 5.47733 & 5.56024 & 5.59341 & 5.62694 & 5.53296 & 5.61800 & 5.65191 & 5.68610\\
    0.4 & 4.98481 & 5.05332 & 5.08021 & 5.10785 & 5.34386 & 5.42207 & 5.45317 & 5.48451 & 5.42763 & 5.50898 & 5.54108 & 5.57324\\
    0.5 & 4.81155 & 4.87494 & 4.89950 & 4.92489 & 5.22228 & 5.29652 & 5.32581 & 5.35531 & 5.33417 & 5.41278 & 5.44327 & 5.47366\\
    0.6 & 4.66432 & 4.72327 & 4.74586 & 4.76938 & 5.11488 & 5.18557 & 5.21331 & 5.24127 & 5.25443 & 5.33106 & 5.36019 & 5.38910\\
    0.7 & 4.53822 & 4.59333 & 4.61424 & 4.63616 & 5.02073 & 5.08810 & 5.11447 & 5.14118 & 5.18772 & 5.26290 & 5.29093 & 5.31861\\
    0.8 & 4.42928 & 4.48107 & 4.50052 & 4.52109 & 4.93780 & 5.00201 & 5.02716 & 5.05280 & 5.13245 & 5.20647 & 5.23366 & 5.26037\\
    0.9 & 4.33445 & 4.38330 & 4.40151 & 4.42090 & 4.86418 & 4.92515 & 4.94915 & 4.97387 & 5.08704 & 5.15984 & 5.18638 & 5.21235\\
    1.0 & 4.25128 & 4.29753 & 4.31466 & 4.33303 & 4.79803 & 4.85574 & 4.87859 & 4.90248 & 5.04996 & 5.12126 & 5.14722 & 5.17264\\
    1.1 & 4.17782 & 4.22181 & 4.23799 & 4.25545 & 4.73806 & 4.79234 & 4.81404 & 4.83716 & 5.02010 & 5.08921 & 5.11451 & 5.13955\\
    1.2 & 4.11255 & 4.15453 & 4.16988 & 4.18653 & 4.68319 & 4.73395 & 4.75444 & 4.77684 & 4.99613 & 5.06247 & 5.08689 & 5.11165\\
    1.3 & 4.05423 & 4.09441 & 4.10902 & 4.12496 & 4.63261 & 4.67979 & 4.69906 & 4.72077 & 4.97718 & 5.04009 & 5.06323 & 5.08774\\
    1.4 & 4.00185 & 4.04042 & 4.05436 & 4.06966 & 4.58567 & 4.62929 & 4.64735 & 4.66840 & 4.96206 & 5.02120 & 5.04264 & 5.06688\\
    1.5 & 3.95458 & 3.99169 & 4.00503 & 4.01975 & 4.54196 & 4.58201 & 4.59889 & 4.61931 & 4.95004 & 5.00520 & 5.02443 & 5.04829\\
    \hline
    \end{tabular*}
    \caption{Final $C_d$ values at $T=30$ for the sedimenting sphere case, using the eigenvalue-free Giesekus formulation at different Weissenberg numbers. Again, three different mobility factors $\alpha \in \{0.1, 0.01, 0.001\}$ were considered.}
    \label{tab:Cd_values_sedimentingSphere_GiesekusLog}
\end{sidewaystable*}
Our data agrees with the literature. We observe a noticeable mesh convergence towards the compared values. Furthermore, increasing Weissenberg numbers result in decreasing $K$ values, which also agrees with the literature. For decreasing $\alpha$ values, an expected convergence of $K$ towards the corresponding values in Tab.~\ref{tab:Cd_values_sedimentingSphere_OBLog} is observed.

\section{Conclusion and Outlook}
In this paper, we have shown how the $f(\opad \mathbf{\Psi})$-based formulation that was first introduced in~\cite{Knechtges2018} can be used to engineer an eigenvalue-free numerical algorithm for the log-conformation formulation.

In the course of our analysis, we first have proven the equivalence of this formulation to many different log-conformation formulations, including the original formulation by Fattal~and~Kupferman~\cite{Fattal2004}.

The new algorithm is in principle not tied to a specific discretization scheme of the resulting constitutive equation. However, in order to verify our algorithm, we have shown a working implementation in the RheoTool~\cite{rheoTool,pimenta2017stabilization} framework, which is based on OpenFOAM\textsuperscript{\textregistered}~\cite{Weller1998}. The resulting implementation was successfully validated on the confined cylinder and sedimenting sphere benchmarks.

For the future, we plan to bridge the performance gap of our prototypical implementation in comparison to the standard implementation by exploiting the matrix symmetries even more. Furthermore, we mostly see the application of this eigenvalue-free algorithm in areas that have so far been hindered by the eigenvalue decomposition. One is certainly bringing more of these heavy computations per finite volume cell onto the GPU. Another is that the algorithm facilitates the development of semi-implicit or fully implicit discretization schemes, in the same vein as ~\cite{Knechtges2014} facilitated adoption of automatic differentiation methods in ~\cite{Zwicke2016}. The latter may allow us to perform more efficient simulations with larger time-steps in the future.

\section{Acknowledgments}
This Project is supported by the Federal Ministry for Economic Affairs and Climate Action (BMWK) on the basis of a decision by the German Bundestag.

In addition, the third author thanks MAGMA Gießereitechnologie GmbH for the freedom to work on cutting-edge research topics.

Last but not least, the authors want to thank the reviewers of the Journal of Non-Newtonian Fluid Mechanics for their valuable comments.

\appendix
\section{Poiseuille Inflow Conditions in the Confined Cylinder Benchmark}
\label{sec:PoiseuilleInflowConditions}
As written in Section~\ref{sec:confinedCylinder}, we want to prescribe a fully developed Poiseuille flow at the inflow of the confined cylinder. This poses the question of whether an easy expression to specify $\mathbf{\Psi}$ exists. For $\bm{\tau}$ it is known that
\begin{align}
    \bm{\tau} &= \begin{pmatrix} \tau_{xx} & \tau_{xy} \\
            \tau_{xy} & 0 \end{pmatrix} \, ,
\end{align}
with $\tau_{xx} = 2\lambda \mu_P \left(\partial_y u_x\right)^2$ and $\tau_{xy} = \mu_P\, \partial_y u_x$. The velocity $\mathbf{u}$ is given by
\begin{align}
    \mathbf{u} &= \begin{pmatrix} \frac{3}{8}\Bar{u}\biggl( 4 - \frac{(y-2R)^2}{R^2}\biggr) \\
            0 \end{pmatrix}\, ,
\end{align}
with mean inflow velocity $\Bar{u}=\SI{1}{m/s}$ in all benchmarks. The coordinate system is centered at the lower left corner of the confined cylinder domain (hence the $-2R$ term), as depicted in Fig.~\ref{fig:confinedCylinder}.

The corresponding conformation tensor is thus given by
\begin{align}
    \mathbf{C} &= \mathbf{1} + \frac{\lambda}{\mu_P} \bm{\tau}
        = \mathbf{1} + \begin{pmatrix} 2 l^2 & l \\ l & 0 \end{pmatrix}\, ,
\end{align}
with $l=\lambda\, \partial_y u_x$.

We claim that $\mathbf{\Psi}$ is given by
\begin{align}
    \mathbf{\Psi} &= \log \mathbf{C} = \frac{1}{2} \begin{pmatrix} 
        p - q l^2/o & - q l/o \\ - q l/o & p + q l^2/o
        \end{pmatrix}\, ,
\end{align}
with
\begin{align}
    o &= \sqrt{l^2(1+l^2)} = \seminorm{l} \sqrt{1+l^2}\\
    p &= \log (1+l^2)\\
    q &= \log \left(1 + 2(l^2 -o)\right) = 2 \arsinh \left(-\seminorm{l}\right)\, .
\end{align}
This is the same formulation as it was used for the actual computations in~\cite{Knechtges2014}. Even though the correct formula was used for computations in~\cite{Knechtges2014}, a small error creeped into the formulas printed in~\cite{Knechtges2014}, which unfortunately omitted factors of $l$ in $\Psi_{xy}$ and $\Psi_{yy}$. With this appendix we want to correct this error.

Coming to the proof, we split $\mathbf{\Psi}$ into two parts: one which encodes the traceful part and one traceless part
\begin{align}
    \mathbf{\Psi} &= \frac{p}{2} \mathbf{1} + \mathbf{B}\, ,
\end{align}
with
\begin{align}
    \mathbf{B} &= \frac{ql}{2o} \begin{pmatrix} -l & -1 \\ -1 & l \end{pmatrix}\, .
\end{align}
Note that the identity matrix $\mathbf{1}$ and $\mathbf{B}$ obviously commute and thus allow us to compute the matrix exponential as two factors
\begin{align}
    \exp \mathbf{\Psi} &= \exp\left(\frac{p}{2}\right) \exp \mathbf{B}\\
        &= \sqrt{1+l^2}\, \exp\mathbf{B}\, .
\end{align}
In order to compute $\exp\mathbf{B}$ it is helpful to see that the following identity holds
\begin{align}
    \mathbf{B}^2 &= \frac{q^2}{4} \mathbf{1}\, .
\end{align}
From this, it follows immediately
\begin{align}
    \mathbf{B}^{2n} &= \left(\frac{q}{2}\right)^{2n}\mathbf{1}\\
    \mathbf{B}^{2n+1} &= \left(\frac{q}{2}\right)^{2n+1} \frac{2}{q} \mathbf{B}\, .
\end{align}
Therefore, we can split the computation of $\exp\mathbf{B}$ into two summands
\begin{align}
    \exp\mathbf{B} &= \sum_{n=0}^\infty \frac{1}{n!} \mathbf{B}^n\\
        &= \sum_{n=0}^\infty \frac{1}{(2n)!} \mathbf{B}^{2n}
                + \sum_{n=0}^\infty \frac{1}{(2n+1)!} \mathbf{B}^{2n+1}\\
        &= \cosh \left(\frac{q}{2}\right) \mathbf{1} + \sinh\left(\frac{q}{2}\right)\frac{2}{q}\mathbf{B}\, .
\end{align}
Together with the identity $\cosh(\arsinh(-\seminorm{l})) = \sqrt{1+l^2}$ it follows
\begin{align}
    \exp\mathbf{B} &= \frac{1}{\sqrt{1+l^2}} \begin{pmatrix} 1+2l^2 & l \\ l & 1\end{pmatrix}\, .
\end{align}
In total we obtain
\begin{align}
    \exp\mathbf{\Psi} &= \begin{pmatrix} 1+2l^2 & l \\ l & 1\end{pmatrix}\, ,
\end{align}
which is what had to be proven.

\bibliographystyle{elsarticle-num}
\bibliography{refs}
\end{document}